\documentclass[11pt,a4paper]{article}
\usepackage{jcappub}

\usepackage{float}
\usepackage{latexsym}
\usepackage{graphicx}
\usepackage{amssymb}
\usepackage{amsmath}
\usepackage{amsfonts}
\usepackage{hyperref}
\usepackage{color}
\usepackage{cool}
\usepackage{subfig}
\usepackage{dcolumn}

\begin{document}

\title{Constraints on interacting dark energy models from Planck 2015 and redshift-space distortion data}

\author[a]{Andr\'e A. Costa,}\emailAdd{alencar@if.usp.br}
\author[b]{Xiao-Dong Xu,}\emailAdd{xiaodong.xu@uct.ac.za}
\author[c]{Bin Wang,}\emailAdd{wang\_b@sjtu.edu.cn}
\author[a]{E. Abdalla}\emailAdd{eabdalla@usp.br}

\affiliation[a]{Instituto de F\'isica, Universidade de S\~ao Paulo, C.P.
  66318, 05315-970, S\~ao Paulo, SP, Brazil}
\affiliation[b]{Department of Mathematics and Applied Mathematics, University of Cape Town, Rondebosch 7701, Cape Town, South Africa}
\affiliation[c]{Department of Physics and Astronomy, Shanghai Jiao Tong University, 200240 Shanghai, China}

\date{\today}

\abstract{We investigate phenomenological interactions between dark matter and dark energy and constrain these models by employing the most recent cosmological data including the cosmic microwave background radiation anisotropies from Planck 2015, Type Ia supernovae, baryon acoustic oscillations, the Hubble constant and redshift-space distortions. We find that the interaction in the dark sector parameterized as an energy transfer from dark matter to dark energy is strongly suppressed by the whole updated cosmological data. On the other hand, an interaction between dark sectors with the energy flow from dark energy to dark matter is proved in better agreement with the available cosmological observations. This coupling between dark sectors is needed to alleviate the coincidence problem.}

\arxivnumber{1605.04138}
\maketitle
\flushbottom

\section{Introduction}
Cosmology is one of the fields of science with the quickest development today, especially concerning observations.
The cosmological data from different observations have been rapidly updating in the past decade, not only in numbers but also in their quality. This allows the derivation of more reliable scientific results and tighter constraints on theoretical models in cosmology. The whole available host of updated cosmological data includes those from geometrical measurements of the universe such as the measurement of the Hubble constant ($H_0$), the luminosity distances of Type Ia supernovae (SNIa) and the pattern of baryon acoustic oscillations (BAO) imprinted in the galaxy distribution, and also the temperature of photons from the cosmic microwave background (CMB) and observations of the large-scale structure (LSS).

Recently, Planck Collaboration released their most recent results on the CMB anisotropies \cite{Adam:2015rua}. These data provide the utmost observations on temperature and polarization of the photons from the last scattering surface at redshift around $z=1090$. Comparing with Planck 2013 data, significant improvements have been made in reducing systematic errors in the new data and the overall level of confidence has been significantly increased.
One of the most notable improvements in the Planck 2015 data set is that its residual systematics in polarization maps have been dramatically reduced compared to 2013, and its agreement to WMAP is within a few tenths of
a percent on angular scales from the dipole to the first acoustic peak \cite{Adam:2015rua}. The 2015 Planck results make important contributions in a variety of theoretical analyses in cosmology and contain smaller uncertainties compared with those determined in 2013 results. However areas that were in tension in 2013 with other astrophysical data sets \cite{Ade:2015xua}, such as the abundance of clusters of galaxies, weak gravitational lensing of galaxies or cosmic shear, distances measured with BAO using Lyman-$\alpha$ forest at high-$z$ and the determination of the Hubble constant with the Hubble Space Telescope (HST) \cite{Riess:2011yx,Riess:2016jrr} have been confirmed to still remain in tension today, although the disagreement is lessened in some cases. If these tensions are not related to systematics, they can point out to physics beyond the standard model.

Another observable, which is worthy of mentioning, is the redshift-space distortion (RSD). It has been measured more and more precise in the past few years. This observable is a key signature to disclose the large scale structure, which is believed as a powerful complementary observation to break the possible degeneracy in cosmological models. This is because the dynamical growth history in the cosmological structure can be distinct even if they undergo similar evolution in the background. A lot of measurements on the RSD have been reported, see for example \cite{Song:2008qt,Samushia:2011cs,Blake:2011rj,Tojeiro:2012rp,Reid:2012sw,Beutler:2012px,Hudson:2012gt,delaTorre:2013rpa,Feix:2015dla}. It is expected that including the large scale structure information by adding the RSD measurements can provide a rich harvest from complementary data sets and obtain tight constrains on theoretical models covering a greater range of cosmology.

In this paper we will employ such great amount of precise new cosmological data to test the interaction models in the universe between dark matter and dark energy. It is well known that our universe is undergoing an accelerated expansion driven by a mysterious dark energy occupying nearly 70\% of the energy content of the universe. The galaxies and other large scale structures distributed in our universe are created by dark matter component which occupies 25\% of the energy budget of the universe. Considering that dark energy  and dark matter dominate the energy content of the universe today, it is  reasonable to assume that these dark components can interact between themselves. A dark matter and dark energy interaction is an attractive theoretical model, since it can allow solutions with a constant ratio between energy densities of dark matter and dark energy at late times, which can help to alleviate the coincidence problem in the concordance cosmological model. For a review on theoretical challenges, cosmological implications and observational signatures on interactions between dark sectors can be referred to \cite{Wang:2016lxa} and references therein.

Since the lack of information on the nature and dynamics of dark matter and dark energy, it is difficult to describe these components from first principles. This makes it hard to describe the interaction between them from a fundamental theory. The interaction between dark sectors is usually described phenomenologically by assuming that the interaction only represents a small correction to the evolution history of the Universe. Similar to how interactions behave in particle physics, one expects the coupling kernel between dark sectors to be a function of the energy
densities involving dark energy, dark matter and of time. In Table \ref{models} we present the phenomenological models that have been commonly considered. These models have been confronted with different observational data sets, such as CMB data on temperature and polarization power spectra from WMAP5 \cite{He:2009pd}, WMAP7 \cite{He:2010im} and Planck 2013 \cite{Costa:2013sva,Salvatelli:2013wra}, together with other different external observational data. Recently the new Planck 2015 data have also been used to constrain one of the phenomenological interacting models with the kernel of the interaction in proportional to the energy density of dark energy only \cite{Murgia:2016ccp}. In addition to the phenomenological models of the interaction between dark sectors, some attempts on describing the coupling in a field theory have been studied, see for example \cite{Costa:2014pba} and other related references in the review \cite{Wang:2016lxa}. Besides the data related to the universe expansion history and CMB, the observational data on the large scale structure such as the RSD data sets have also been employed in constraining the interaction models. In combining with the Planck 2013, it was argued that the RSD data can rule out a large interaction rate in $1\sigma$ region \cite{Yang:2014gza,Li:2015vla}. More complete references on testing the interaction models between dark sectors with different observational data sets can be found in the recent review \cite{Wang:2016lxa}.

The main motivation of the present paper is to confront the phenomenological interacting dark energy models to a whole host of updated cosmological data, including the Planck 2015 result, the new RSD data together with other different external data sets. We are going to check the consistency of the constraints on the model parameters obtained from Planck 2013 results and examine the effectiveness of increasing the confidence level and tightening the constraints on model parameters by including the complementary observables such as RSD and other external data sets. We hope that the updated precise data can help to improve limits on the interaction between dark sectors.

Weak lensing is also sensitive to the amount of dark matter and could be used to constrain coupled dark energy models. However, in the current state-of-art, weak lensing measurement has a number of its own problems, therefore one cannot yet draw conclusive results from lensing observations. There are some issues concerning their systematic uncertainties in both their measurement and the modelling of physics (such as intrinsic alignment, baryonic effects on
density evolution, nonlinear evolution of the dark matter field) that need to be better understood \cite{Kilbinger:2012qz,Heymans:2013fya,Joudaki:2016mvz}. Despite of those problems, there are some efforts to constrain coupled dark energy models using weak lensing \cite{Yang:2016evp}. In this work we will not consider weak lensing data.

The organization of the paper is as follows. In section \ref{sec:coupled_models}, we review the evolution description of the background dynamics and the linear perturbation when there is interaction between dark sectors. In section \ref{sec:analysis}, we introduce the observational data that we are going to use and the methods for data analysis. In the following section we report the main results by confronting our models to observational data and we discuss and compare with the results obtained in previous works. Finally, in section \ref{sec:conclusions} we present our conclusions and discussions.

\section{Phenomenological interacting dark energy models}
\label{sec:coupled_models}
We consider a cosmological model with an interaction between dark matter and dark energy. In this model, the conservation of the energy-momentum tensor for dark matter or dark energy satisfies respectively
\begin{equation}\label{EE_int}
\nabla _{\mu} T^{\mu\nu}_{(l)} = Q^{\nu}_{(l)}\,,
\end{equation}
where $(l)$ represents either dark matter with the subscript $(c)$, or dark energy with the subscript $(d)$. The presence of the term $Q^{\nu}_{(l)}$ implies that these components are not conserved independently, and there is an energy-momentum flux between them. However, the Bianchi identity requires that the energy-momentum tensor of the total dark sectors still satisfy the conservation law, such that $Q^{\nu}_{(c)} = -Q^{\nu}_{(d)}$.

We assume that the universe is described by a flat Friedmann-Lemaitre-Robertson-Walker (FLRW) metric with small perturbations over a smooth background. Thus, the line element is given by
\begin{equation}\label{dse}
ds^2 = a^2\left[-(1 + 2\psi)d\eta^2 + 2\partial_i B d\eta dx^i + (1 + 2\phi)\delta_{ij}dx^idx^j + D_{ij}Edx^idx^j\right],
\end{equation}
where
\begin{equation}
D_{ij} = \left(\partial_i\partial_j - \frac{1}{3}\delta_{ij}\nabla^2\right).
\end{equation}
The function $a=a(\eta)$ is the scale factor of the universe and $\eta$ is the conformal time. $\psi$, $B$, $\phi$ and $E$ are functions of space and time describing small deviations from the homogeneous and isotropic universe. Actually, we are only considering scalar perturbations, but, at the linear perturbation level, all other modes decouple and scalar perturbations are responsible for the structure formation.

The matter content is identified with the energy-momentum tensor of a perfect fluid,
\begin{equation}\label{T}
T^{\mu\nu}(\eta,x,y,z) = (\rho + P)U^\mu U^\nu + Pg^{\mu\nu},
\end{equation}
where, for each species, the energy density is written as $\rho(\eta,x,y,z) = \rho(\eta)[1 + \delta(\eta,x,y,z)]$, the pressure is $P(\eta,x,y,z) = P(\eta) + \delta P(\eta,x,y,z)$ and the four-velocity vector is $U^\mu = a^{-1}(1 - \psi, \vec{v}_{(l)})$. Thus, we have split the components of the energy-momentum tensor into background quantities and small perturbations. Combining the conservation equation (\ref{EE_int}) with the line element (\ref{dse}) and the energy-momentum tensor (\ref{T}), we obtain, at the background level, the continuity equations
\begin{alignat}{2}\label{eq:intera_fen}
\dot{\rho}_{c}+3\mathcal{H}\rho_{c}= & a^2Q^{0}_{c}= & +aQ\,,\nonumber \\
\dot{\rho}_{d}+3\mathcal{H}\left(1+\omega\right)\rho_{d}= & a^2Q^{0}_{d}= & -aQ\,.
\end{alignat}
In these equations, $\mathcal{H}$ is the Hubble parameter expressed in conformal time, $\mathcal{H} \equiv \dot{a}/a = a H$, and the dot represents the derivative with respect to the conformal time. $\omega = P_d/\rho_d$ is the equation of state of dark energy and $Q$ is the energy transfer in cosmic time coordinates, which will be written as $Q=3H(\lambda_1\rho_c + \lambda_2\rho_d)$. Table \ref{models} shows the phenomenological models we are going to  investigate. The inequalities are the stability conditions for the models \cite{He:2008si,Gavela:2009cy}.
\begin{table}[htb!]
\centering
\caption{In this table we present different stable phenomenological interacting dark energy models.}
\begin{tabular}{|c|c|c|c|}
\hline
    Model & Q & DE EoS & Constraints \\
    \hline
    I & $3\lambda_{2}  H\rho _{d}$ & $-1 <  \omega < 0$ & $\lambda_{2} < 0$ \\
    \hline
    II & $3\lambda_{2} H\rho_{d}$  & $\omega < -1$ & $0 < \lambda_{2} < -2\omega \Omega_{c}$ \\
\hline
    III & $3\lambda_{1} H\rho_{c}$  & $\omega < -1$ & $0 < \lambda_{1} < -\omega /4$ \\
    \hline
    IV & $3\lambda H \left(\rho_{d} + \rho_{c} \right) $ & $\omega < -1$ & $0 < \lambda < -\omega /4$ \\
\hline
\end{tabular}
\label{models}
\end{table}

Taking into account the first-order perturbation equations, the energy-momentum conservation in the synchronous gauge yields \cite{Costa:2013sva}
\begin{align}
     \dot{\delta}_{c} & =  -(kv_{c} + \frac{\dot{h}}{2}) + 3\mathcal{H}\lambda_{2} \frac{1}{r} \left( \delta_{d}-\delta_{c} \right)\,, \label{linear_pert_1} \\
     \dot{\delta}_{d} & = -\left(1+\omega \right) (k v_{d} + \frac{\dot{h}}{2})+3\mathcal{H} (\omega - c_{e}^{2}) \delta_{d}+3\mathcal{H} \lambda_{1} r \left( \delta _{d}-\delta _{c} \right)  \nonumber \\
                            & -3\mathcal{H} \left( c_{e}^{2}-c_{a}^{2} \right) \left[ 3 \mathcal{H} \left( 1+\omega \right) + 3\mathcal{H} \left( \lambda_{1} r+\lambda_{2} \right) \right]\frac{v_{d}}{k}      \,, \label{linear_pert_2} \\
     \dot{v}_{c} & = -\mathcal{H}v_{c} -3\mathcal{H}(\lambda_{1} + \frac{1}{r}\lambda_{2})v_{c} \,, \label{linear_pert_3} \\
     \dot{v}_{d} & = -\mathcal{H}\left(1-3c_{e}^{2} \right) v_{d}+\frac{3\mathcal{H}}{1+\omega} \left( 1+c_{e}^{2} \right) \left(\lambda_{1} r+\lambda_{2} \right) v_{d}+ \frac{kc_{e}^{2}\delta _{d}}{1+\omega}\,,
     \label{linear_pert_4}
\end{align}
where $v_{(l)}$ is the peculiar velocity of the $(l)$ component and $h = 6\phi$ is the synchronous gauge metric perturbation. We have defined $r \equiv \rho_c/\rho_d$, $c_e$ is the effective sound speed and $c_a$ is the adiabatic sound speed for the dark energy fluid at the rest frame.

\section{Cosmological data sets}
\label{sec:analysis}
In order to confront our interacting models described in the previous section to the observational data, we will employ the Bayesian statistics. We implement the background dynamics and linear perturbation equations of the theoretical model into the CAMB code \cite{Lewis:1999bs} and derive the theoretical predictions. On the other hand, we use the CosmoMC code \cite{Lewis:2002ah,Lewis:2013hha} to estimate the parameters that best describe the observational data.

We use the most recent results of CMB anisotropies from Planck 2015 \cite{Adam:2015rua}. In our analysis we take the data for the low-$l$ ($l = 2-29$) temperature and polarization spectrum (TT, TE, EE, BB), combined with the high-$l$ TT, TE and EE CMB data in the range $l = 30-2508$ for TT and $l = 30-1996$ for TE and EE. In addition to the CMB data, we also consider BAO measurements. We combine four different BAO measurements: the 6dFGS at effective redshift $z_{eff} = 0.106$ \cite{Beutler:2011hx}, the SDSS-MGS at effective redshift $z_{eff} = 0.15$ \cite{Ross:2014qpa}, the BOSS-LOWZ at effective redshift $z_{eff} = 0.32$ and the CMASS-DR11 at effective redshift $z_{eff} = 0.57$ \cite{Anderson:2013zyy}. We further employ Type Ia supernovae data to better constrain the parameters of our interacting models. For the SNIa data we consider the ``Joint Light-curve Analysis'' (JLA) \cite{Betoule:2014frx}, which is a combination of SNLS and SDSS together with several samples of low redshift supernovae. Furthermore, we include a conservative gaussian prior for $H_0$ based in recent results \cite{Efstathiou:2013via}
\begin{equation}
H_0 = 70.6 \pm 3.3 \, \mathrm{ km \, s^{-1}Mpc^{-1}}.
\end{equation}

The distances to galaxies are usually inferred through their redshifts. This induces an error on the distances since the redshift is also affected by the peculiar motions of galaxies. Thus, such peculiar velocities produce anisotropies in the transverse versus line-of-sight directions in the redshift space. On large scales, the galaxies tend to fall towards concentrations, therefore, the velocity field is coupled to the density field. This generates a systematic effect in the  redshift space that can be used to constrain the growth rate of structure. Following this idea, several groups have worked to constrain the parameter combination $f\sigma_8(z)$ which is considered to be model independent, where
\begin{equation}\label{f}
f \equiv \frac{d \ln{\delta_m}}{d \ln{a}} = \frac{\mathcal{H}^{-1}}{\delta_m}\left(\frac{\dot{\delta}_c \rho_c + \dot{\delta}_b \rho_b}{\rho_m} + \delta_c\frac{aQ}{\rho_m} - \delta_m\frac{aQ}{\rho_m}\right) .
\end{equation}
The indices $m$, $b$ represent the total matter (excluding neutrinos) and baryons, respectively.  The last equality shows the dependence of the growth rate on the interaction between dark sectors, which comes from the fact that cold dark matter no longer evolves in the same way as baryons in an interacting scenario. If the interaction is null, we re-obtain the standard result.

\begin{figure}[h!]
\subfloat[$\Lambda$CDM]{ \includegraphics[width=0.45\textwidth]{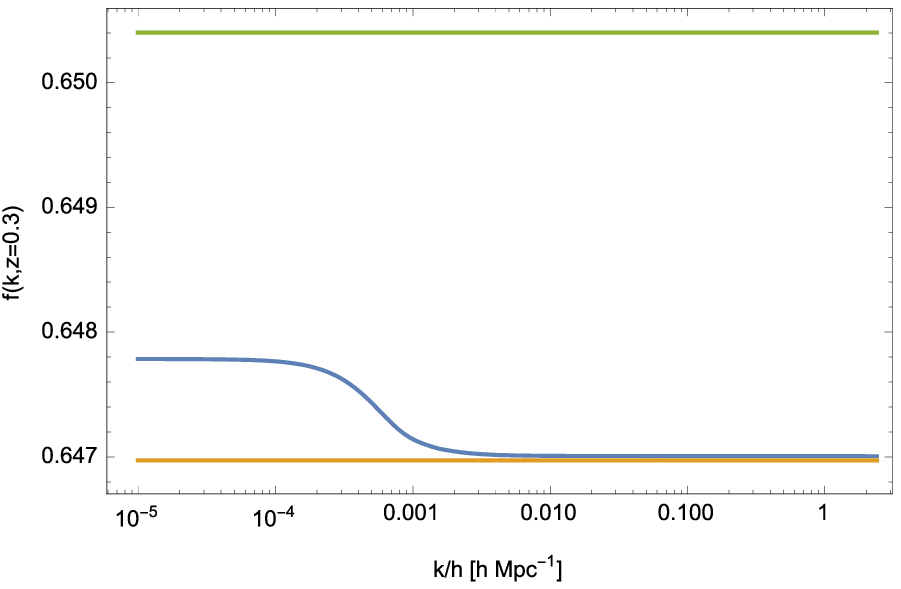}}
\subfloat[Interacting Model]{ \includegraphics[width=0.45\textwidth]{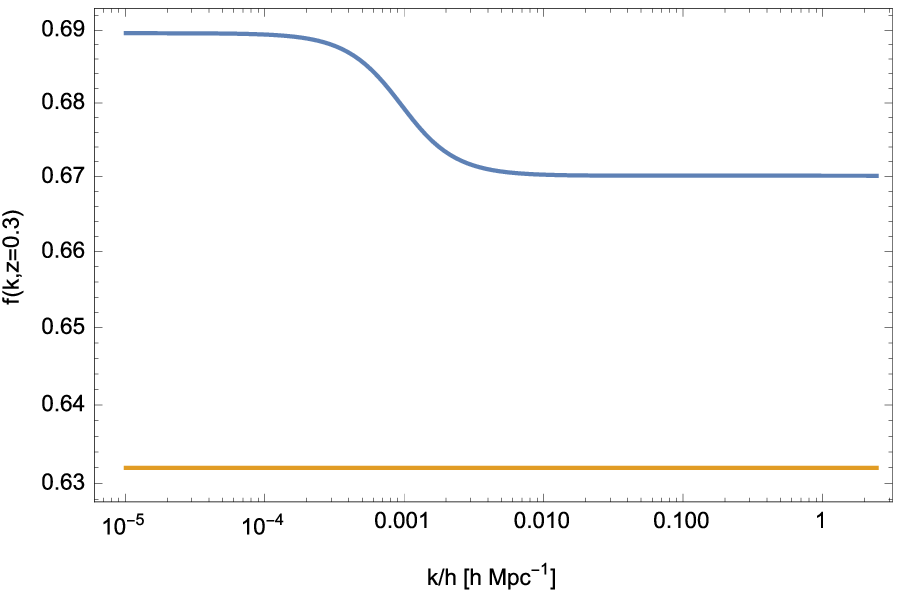}}
\caption{The growth rate $f$ as a function of the wave number $k$. The blue line is for $f = \frac{d\ln{\delta_m}}{d\ln{a}}$, the orange line is for $f= \left(\frac{\sigma_{vd}}{\sigma_8}\right)^2$ and the green line is for $f= \Omega_m^{0.545}$.}
\label{fofk}
\end{figure}
The growth rate defined in equation (\ref{f}) actually is a function of the wave number $k$ as well as the redshift, $f = f(k,z)$. We plot the dependence of $f$ with $k$ at a fixed redshift in Fig. \ref{fofk}. Since matter structures grow on spatial scales much smaller than that of the Hubble horizon $(k \gg \mathcal{H})$, we can take the subhorizon condition in the calculation of $f$ by fixing $k$ at a large enough value. We can see from Fig. \ref{fofk} that $f(k)$ is approximately constant for several orders of magnitude from the largest $k$, a pattern that we can observe in all the redshifts of interest. Therefore, that choice for $k$ seems reasonable.

\begin{table*}[h!]
\caption{RSD data} \centering \label{RSD}
\begin{tabular}{ccccc}
\hline
\multicolumn{1}{c}{z} & \multicolumn{1}{c}{f$\sigma_8(z)$} & \multicolumn{1}{c}{Reference} \\
\hline
0.02	 & 0.360 $\pm$	0.040 & \cite{Hudson:2012gt} \\
0.067 & 	0.423 $\pm$	0.055 & \cite{Beutler:2012px} \\
0.10	 & 0.37 $\pm$	0.13 & \cite{Feix:2015dla} \\
0.17	 & 0.51 $\pm$	0.06  & \cite{Song:2008qt} \\
0.22	 & 0.42 $\pm$	0.07 & \cite{Blake:2011rj} \\
0.25	 & 0.3512 $\pm$	0.0583 & \cite{Samushia:2011cs} \\
0.30	 & 0.407 $\pm$	0.055 & \cite{Tojeiro:2012rp} \\
0.35	 & 0.440 $\pm$	0.050 & \cite{Song:2008qt} \\
0.37	 & 0.4602 $\pm$	0.0378 & \cite{Samushia:2011cs} \\
0.40	 & 0.419 $\pm$	0.041 & \cite{Tojeiro:2012rp} \\
0.41	 & 0.45 $\pm$	0.04 & \cite{Blake:2011rj} \\
0.50	 & 0.427 $\pm$	0.043 & \cite{Tojeiro:2012rp} \\
0.57	 & 0.427 $\pm$	0.066 & \cite{Reid:2012sw} \\
0.6	 & 0.43 $\pm$	0.04 & \cite{Blake:2011rj} \\
0.6	& 0.433 $\pm$	0.067 & \cite{Tojeiro:2012rp} \\
0.77	 & 0.490 $\pm$	0.180 & \cite{Song:2008qt} \\
0.78	 & 0.38 $\pm$	0.04 & \cite{Blake:2011rj} \\
0.80	 & 0.47 $\pm$	0.08 & \cite{delaTorre:2013rpa} \\
\hline
\end{tabular}
\end{table*}
Another feature we learn from Fig. \ref{fofk} is that $f$ defined in Eq. (\ref{f}), agrees very well with $f \equiv [\sigma_8^{(vd)}(z)]^2/[\sigma_8^{(dd)}(z)]^2$ for the $\Lambda$CDM model \cite{Ade:2015xua}, where $\sigma_8^{(vd)}$ measures the smoothed density-velocity correlation and is defined analogously to $\sigma_8 \equiv \sigma_8^{(dd)}$. However, for the interacting models, the difference between these two definitions can lead to differences on $f\sigma_8(z)$ of the same order or more as the error on the observational measurements reported in Table \ref{RSD}.

\begin{figure}[h!]
\subfloat[$f(z)$]{
\includegraphics[width=0.45\textwidth]{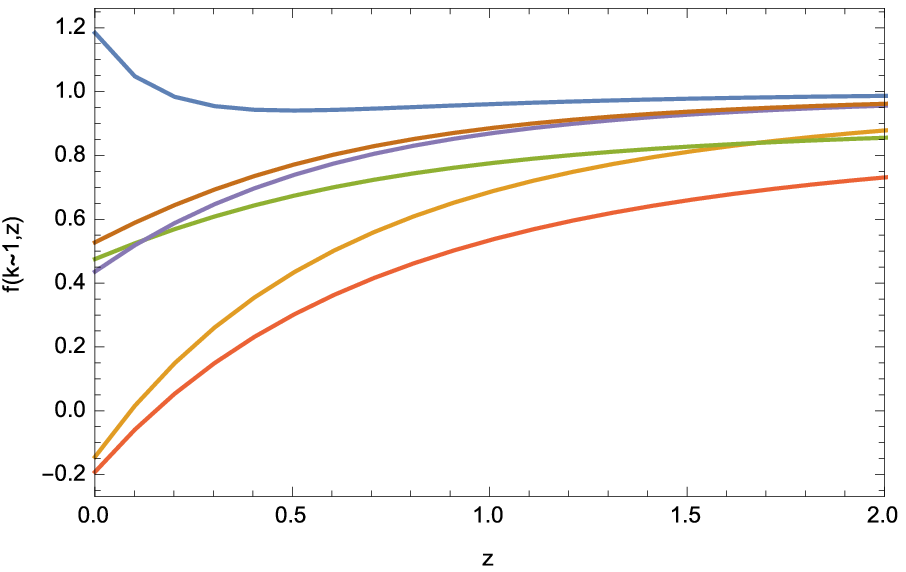}\label{fofz}}
\subfloat[$f\sigma_8(z)$]{
\includegraphics[width=0.45\textwidth]{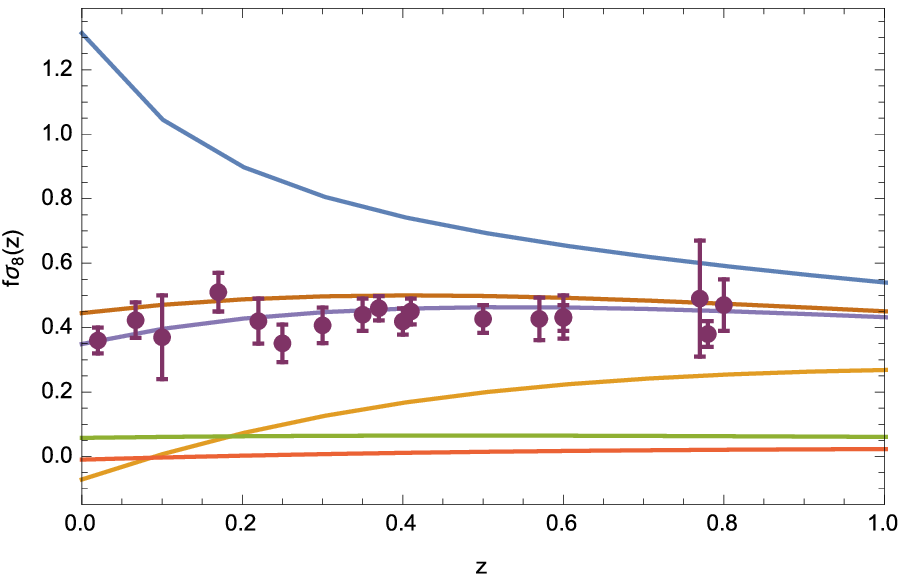}\label{fsigma8}}
\caption{Evolutions of $f$ and $f\sigma_8(z)$ as a function of redshift.  The blue line is for Model I, the orange line for Model II, the green line for Model III and the red line is for Model IV.  All lines are plotted with the same strength of the interaction, $\lambda_i= 0.1$, where the equations of state of dark energy are taken as $\omega \approx -1$. The purple line is the best fit for Model II, as described in the sixth column of Table \ref{tab.model2_RSD}, and the brown line is plotted with the same parameters, but without the interaction. We  include the data in Table \ref{RSD} with corresponding error bars in the right panel.}
\end{figure}

In Fig. \ref{fofz} we show the evolutions of $f$ for our interacting models. We notice that the growth factor of interacting dark energy models can behave in very different ways as compared to the $\Lambda$CDM model. For Model I, the growth factor $f$ can be enhanced at the present moment. While for Model II and IV, the growth factor can be a negative value   at the present if the coupling is strong enough.

In order to confront our models to large scale structure observations, in Table \ref{RSD} we list the available $f\sigma_8(z)$ data sets at different redshifts. Note that the measurement of $f\sigma_8$ at $z=0.02$ \cite{Hudson:2012gt} is not obtained using RSD observations, but is inferred from the peculiar velocities directly from the distance measurements.

To confront the interacting models to observational data, we need to carry out numerical fitting analysis and we need to set the priors of cosmological parameters as listed in Table \ref{parameters}. Furthermore, we fix the relativistic number of degrees of freedom to $N_{eff} = 3.046$, the total neutrino mass to $\sum m_\nu = 0.06 \, \mathrm{eV}$ and the spectrum lensing normalization to $A_L = 1$. We also use a big bang nucleosynthesis consistent scenario to predict the primordial helium abundance. Finally, we set the statistical convergence according to the Gelman and Rubin criterion $R -1 = 0.03$ \cite{Gelman:1992zz}.
\begin{table}[h!]
\centering \caption{The priors for cosmological
parameters considered in the analysis for
different interaction models.}
\begin{tabular}{|c|c|}
\hline
Parameters & Prior \\
\hline
$\Omega_b h^2$ & $[0.005, 0.1] $\\
\hline
$\Omega_c h^2$ & $[0.001, 0.99]$ \\
\hline
$100\theta$ & $[0.5, 10]$ \\
\hline
$\tau$ & $[0.01, 0.8]$ \\
\hline
$n_s$ & $[0.9, 1.1]$ \\
\hline
$\log(10^{10} A_s)$ & $[2.7, 4]$ \\
\hline
&
\begin{tabular}{p{1,7cm}|p{1,7cm}|p{1,7cm}|p{1,7cm}}
Model I & Model II & Model III & Model IV
\end{tabular}
\\
\hline
$\omega$ &
\begin{tabular}{p{1,7cm}|p{1,7cm}|p{1,7cm}|p{1,7cm}}
$[-1, -0.3]$ & $[-3, -1]$ & $[-3, -1]$ & $[-3, -1]$
\end{tabular}
\\
\hline
$\lambda$ &
\begin{tabular}{p{1,7cm}|p{1,7cm}|p{1,7cm}|p{1,7cm}}
$[-0.4, 0]$ & $[0, 0.4]$ & $[0, 0.01]$ & $[0, 0.01]$
\end{tabular}
\\
\hline
\end{tabular}
\label{parameters}
\end{table}

\section{Numerical fitting results}
\label{sec:results}
In our numerical analysis we explore different combinations of the observational data sets. We first report the results by using only the CMB data sets from Planck 2015 and then combine it with other data sets such as BAO, SNIa and finally test the interacting model with the combination of all of these data together with the $H_0$ data.

\begin{figure}[htb!]
\subfloat[Model I]{
\includegraphics[width=0.85\textwidth]{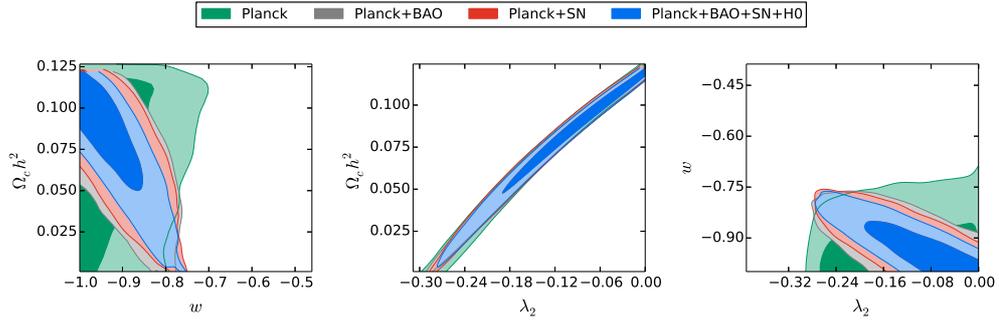}\label{2d_dist1}}\\
\subfloat[Model II]{
\includegraphics[width=0.85\textwidth]{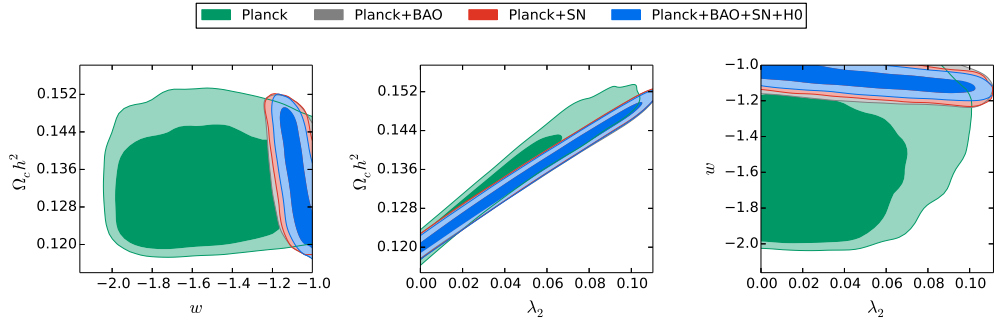}\label{2d_dist2}}\\
\subfloat[Model III]{
\includegraphics[width=0.85\textwidth]{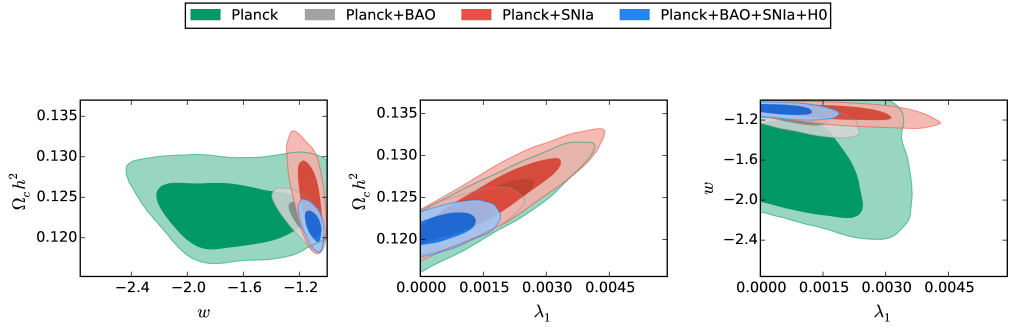}\label{2d_dist3}}\\
\subfloat[Model IV]{
\includegraphics[width=0.85\textwidth]{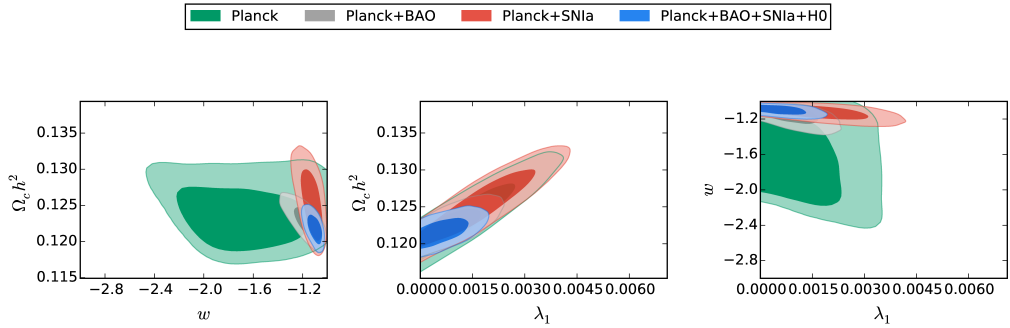}\label{2d_dist4}}
\caption{2-D distribution for selected parameters.}
\end{figure}

\begin{table*}[h!]
\caption{Cosmological parameters - Model I.} \centering \label{tab.model1}
\resizebox{\textwidth}{!}{
\begin{tabular}{lllllllll}
\hline
    & \multicolumn{2}{c}{Planck} & \multicolumn{2}{c}{Planck+BAO} & \multicolumn{2}{c}{Planck+SNIa} & \multicolumn{2}{c}{Planck+BAO+SNIa+H0} \\
    \cline{2-9}
    Parameter & Best fit  & 68\% limits & Best fit  & 68\% limits & Best fit  & 68\% limits & Best fit  & 68\% limits \\
    \hline
$\Omega_b h^2$ & 0.02231 & $0.0222^{+0.00016}_{-0.00016}$ & 0.02213 & $0.02223^{+0.000154}_{-0.000153}$ & 0.02227 & $0.0222^{+0.000159}_{-0.000158}$ & 0.02224 & $0.02223^{+0.000159}_{-0.000158}$\\
$\Omega_c h^2$ & 0.04788 & $0.07131^{+0.0472}_{-0.024}$ & 0.1085 & $0.078^{+0.0365}_{-0.017}$ & 0.09446 & $0.0785^{+0.0348}_{-0.0166}$ & 0.08725 & $0.0792^{+0.0348}_{-0.0166}$\\
$100\theta_{MC}$ & 1.045 & $1.044^{+0.0015}_{-0.00329}$ & 1.042 & $1.043^{+0.00102}_{-0.00242}$ & 1.042 & $1.043^{+0.000996}_{-0.00234}$ & 1.043 & $1.043^{+0.000996}_{-0.00234}$\\
$\tau$ & 0.08204 & $0.08063^{+0.0171}_{-0.0169}$ & 0.07242 & $0.08214^{+0.0171}_{-0.0171}$ & 0.102 & $0.08041^{+0.0167}_{-0.0168}$ & 0.09792 & $0.08204^{+0.0167}_{-0.0168}$\\
${\rm{ln}}(10^{10} A_s)$ & 3.102 & $3.097^{+0.0328}_{-0.0329}$ & 3.079 & $3.099^{+0.033}_{-0.0334}$ & 3.137 & $3.096^{+0.0324}_{-0.0325}$ & 3.131 & $3.099^{+0.0324}_{-0.0325}$\\
$n_s$ & 0.9639 & $0.9633^{+0.00472}_{-0.00514}$ & 0.9649 & $0.9646^{+0.00454}_{-0.0046}$ & 0.9634 & $0.9638^{+0.00485}_{-0.00479}$ & 0.9658 & $0.9645^{+0.00485}_{-0.00479}$\\
$w$ & -0.9765 & $-0.9031^{+0.023}_{-0.0959}$ & -0.9977 & $-0.9124^{+0.0235}_{-0.0866}$ & -0.9787 & $-0.9151^{+0.0222}_{-0.0839}$ & -0.9434 & $-0.9191^{+0.0222}_{-0.0839}$\\
$\lambda_2$ & -0.1831 & $-0.1297^{+0.13}_{-0.0448}$ & -0.03784 & $-0.1137^{+0.0943}_{-0.0481}$ & -0.07739 & $-0.1141^{+0.085}_{-0.0506}$ & -0.09291 & $-0.1107^{+0.085}_{-0.0506}$\\
\hline
$H_0$ & 72.36 & $68.1^{+3.99}_{-3.2}$ & 67.95 & $68.05^{+1.29}_{-1.55}$ & 68.68 & $67.99^{+1.43}_{-1.44}$ & 68.45 & $68.18^{+1.43}_{-1.44}$\\
$\Omega_{de}$ & 0.8647 & $0.7899^{+0.0932}_{-0.106}$ & 0.7156 & $0.7806^{+0.0453}_{-0.0841}$ & 0.7511 & $0.7795^{+0.0454}_{-0.0797}$ & 0.7649 & $0.7796^{+0.0454}_{-0.0797}$\\
$\Omega_m$ & 0.1353 & $0.2101^{+0.106}_{-0.0926}$ & 0.2844 & $0.2194^{+0.0841}_{-0.0453}$ & 0.2489 & $0.2205^{+0.0797}_{-0.0454}$ & 0.2351 & $0.2204^{+0.0797}_{-0.0454}$\\
$\sigma_8$ & 1.622 & $1.438^{+0.143}_{-0.789}$ & 0.9007 & $1.244^{+0.0921}_{-0.478}$ & 1.024 & $1.233^{+0.087}_{-0.446}$ & 1.059 & $1.219^{+0.087}_{-0.446}$\\
${\rm{Age}}/{\rm{Gyr}}$ & 13.71 & $13.81^{+0.058}_{-0.0916}$ & 13.81 & $13.8^{+0.0305}_{-0.0306}$ & 13.78 & $13.8^{+0.0351}_{-0.035}$ & 13.79 & $13.8^{+0.0351}_{-0.035}$
\end{tabular}
}
\bigskip
\resizebox{\textwidth}{!}{
\begin{tabular}{ccccc}
\hline
$\chi^2_{min} =$    & $\chi^2_{CMB} + \chi^2_{prior}$ & $\chi^2_{CMB} + \chi^2_{prior} + \chi^2_{BAO}$ & $\chi^2_{CMB} + \chi^2_{prior} + \chi^2_{SNIa}$ & $\chi^2_{CMB} + \chi^2_{prior} + \chi^2_{BAO} + \chi^2_{SNIa} + \chi^2_{H_0}$ \\
\cline{2-5}
     & 12935.29 + 10.89 & 12936.04 + 9.76 + 5.20 & 12935.22 + 12.32 + 696.44 & 12939.85 + 8.89 + 4.93 + 695.39 + 0.44 \\
\hline
\end{tabular}
}
\end{table*}
\begin{figure}[h!]
\subfloat[Model I]{
\includegraphics[width=\textwidth]{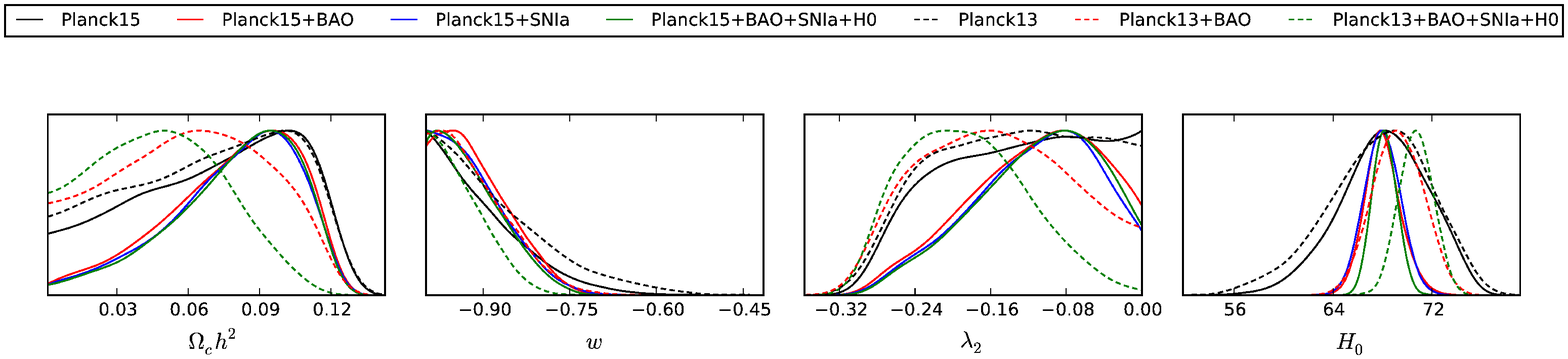}\label{1d_dist1}}\\
\subfloat[Model II]{
\includegraphics[width=\textwidth]{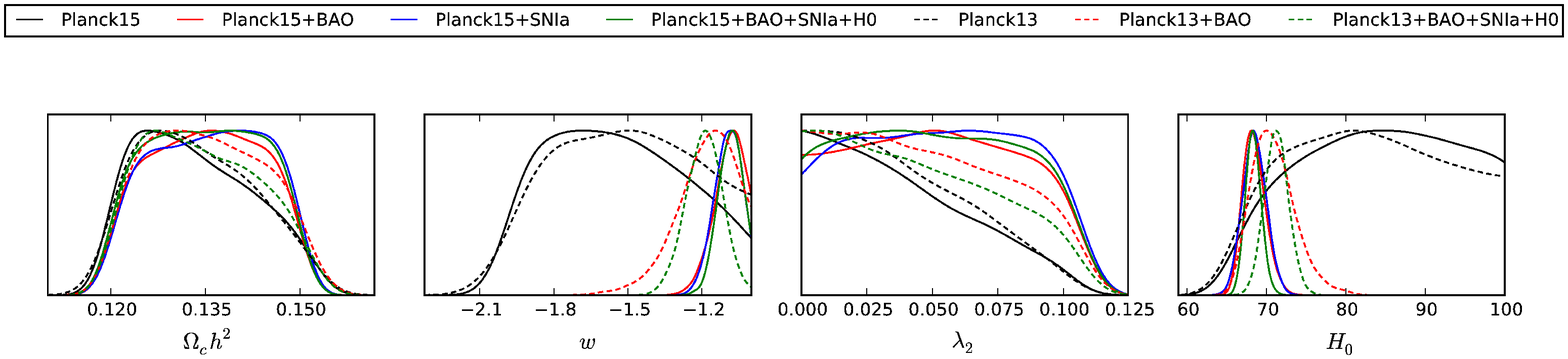}\label{1d_dist2}}\\
\subfloat[Model III]{
\includegraphics[width=\textwidth]{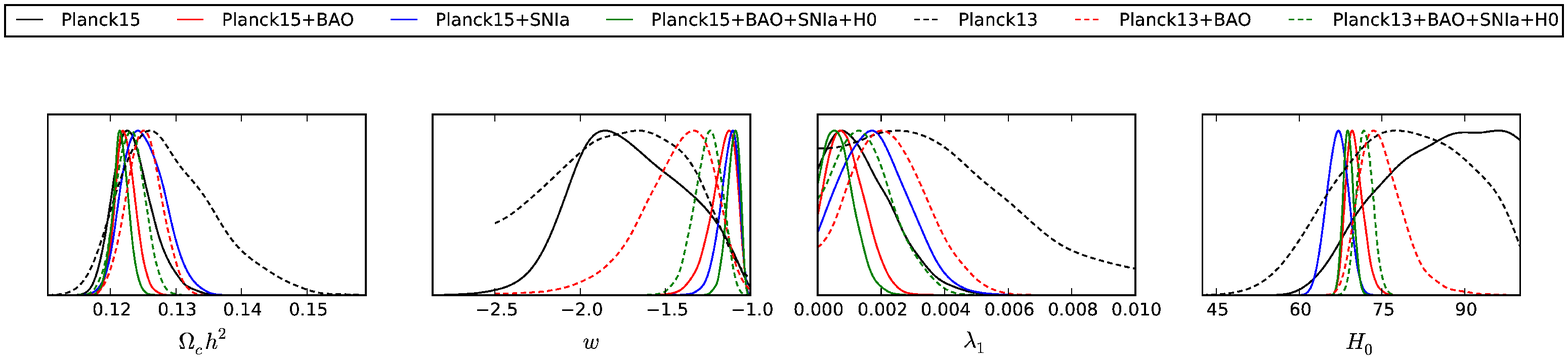}\label{1d_dist3}}\\
\subfloat[Model IV]{
\includegraphics[width=\textwidth]{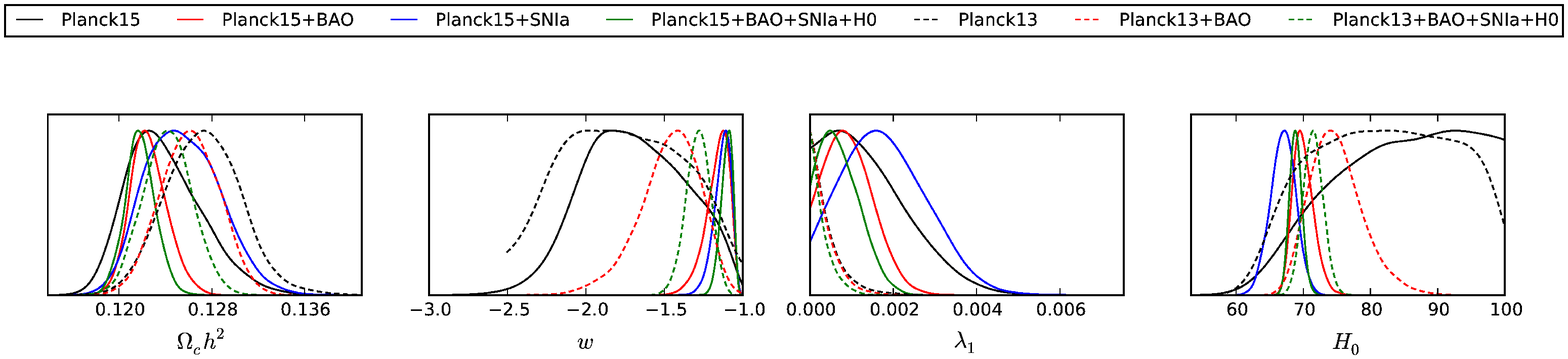}\label{1d_dist4}}
\caption{1-D distribution for selected parameters.}
\end{figure}
In Table \ref{tab.model1} we list the best fit and 68\% C.L. values for relevant parameters of Model I from different analyses. The 1-D marginalized posterior distribution is shown in Fig. \ref{1d_dist1} for some parameters of interest, where we also include our previous results in \cite{Costa:2013sva} by using Planck 2013 results for better comparison. Figure \ref{2d_dist1} have some 2-D posterior distributions. From those results, we observe that Planck 2015 data alone produce a little improvement in the constraints, but without any significant difference from our previous results by using Planck 2013 results. However, when we take into account the joint constraints with low redshift measurements, there are some deviations. In particular, the joint analysis with the old data present a preference for a smaller value of $\Omega_c h^2$ than employing the current data. The $1\sigma$ range for the coupling constant is more consistent from different new data sets compared with that in \cite{Costa:2013sva}. In Fig. \ref{1d_dist1} we see that the interaction parameter $\lambda_2$ becomes less negative by using the combined new data sets.

\begin{table*}[h!]
\caption{Cosmological parameters - Model II.}\centering \label{tab.model2}
\resizebox{\textwidth}{!}{
\begin{tabular}{lllllllll}
\hline
    & \multicolumn{2}{c}{Planck} & \multicolumn{2}{c}{Planck+BAO} & \multicolumn{2}{c}{Planck+SNIa} & \multicolumn{2}{c}{Planck+BAO+SNIa+H0} \\
    \cline{2-9}
    Parameter & Best fit  & 68\% limits & Best fit  & 68\% limits & Best fit  & 68\% limits & Best fit  & 68\% limits \\
    \hline
$\Omega_b h^2$ & 0.02232 & $0.02225^{+0.000162}_{-0.000161}$ & 0.02221 & $0.02223^{+0.000149}_{-0.000148}$ & 0.02217 & $0.02222^{+0.000152}_{-0.000155}$ & 0.02229 & $0.02224^{+0.000152}_{-0.000155}$\\
$\Omega_c h^2$ & 0.1314 & $0.1334^{+0.00692}_{-0.0125}$ & 0.1405 & $0.1352^{+0.00958}_{-0.00972}$ & 0.1436 & $0.1357^{+0.0111}_{-0.00861}$ & 0.1314 & $0.1351^{+0.0111}_{-0.00861}$\\
$100\theta_{MC}$ & 1.04 & $1.04^{+0.000651}_{-0.000562}$ & 1.04 & $1.04^{+0.00056}_{-0.000556}$ & 1.039 & $1.04^{+0.000541}_{-0.000595}$ & 1.04 & $1.04^{+0.000541}_{-0.000595}$\\
$\tau$ & 0.07543 & $0.07653^{+0.0177}_{-0.0174}$ & 0.08067 & $0.08071^{+0.0172}_{-0.0168}$ & 0.08452 & $0.07923^{+0.0174}_{-0.0173}$ & 0.09871 & $0.081^{+0.0174}_{-0.0173}$\\
${\rm{ln}}(10^{10} A_s)$ & 3.082 & $3.088^{+0.0342}_{-0.0337}$ & 3.099 & $3.096^{+0.0334}_{-0.0328}$ & 3.104 & $3.093^{+0.0337}_{-0.0331}$ & 3.131 & $3.097^{+0.0337}_{-0.0331}$\\
$n_s$ & 0.9657 & $0.9638^{+0.00477}_{-0.00475}$ & 0.9664 & $0.9641^{+0.00441}_{-0.00448}$ & 0.9653 & $0.9636^{+0.00477}_{-0.00476}$ & 0.9629 & $0.9643^{+0.00477}_{-0.00476}$\\
$w$ & -1.872 & $-1.55^{+0.235}_{-0.358}$ & -1.131 & $-1.094^{+0.08}_{-0.0351}$ & -1.133 & $-1.097^{+0.0651}_{-0.0448}$ & -1.087 & $-1.088^{+0.0651}_{-0.0448}$\\
$\lambda_2$ & 0.02931 & $0.03884^{+0.0116}_{-0.0388}$ & 0.07053 & $0.05237^{+0.0287}_{-0.0407}$ & 0.08191 & $0.0538^{+0.0349}_{-0.0355}$ & 0.03798 & $0.05219^{+0.0349}_{-0.0355}$\\
\hline
$H_0$ & 96.2 & $83.88^{+13.3}_{-7.86}$ & 69.15 & $68.47^{+1.19}_{-1.63}$ & 68.66 & $68.49^{+1.47}_{-1.46}$ & 68.76 & $68.35^{+1.47}_{-1.46}$\\
$\Omega_{de}$ & 0.8331 & $0.7688^{+0.0778}_{-0.0353}$ & 0.6583 & $0.6622^{+0.0276}_{-0.0243}$ & 0.647 & $0.6613^{+0.0277}_{-0.0243}$ & 0.6735 & $0.6616^{+0.0277}_{-0.0243}$\\
$\Omega_m$ & 0.1669 & $0.2312^{+0.0353}_{-0.0778}$ & 0.3417 & $0.3378^{+0.0243}_{-0.0276}$ & 0.353 & $0.3387^{+0.0243}_{-0.0277}$ & 0.3265 & $0.3384^{+0.0243}_{-0.0277}$\\
$\sigma_8$ & 0.9852 & $0.9016^{+0.0945}_{-0.094}$ & 0.7616 & $0.7792^{+0.0394}_{-0.049}$ & 0.7513 & $0.7773^{+0.0349}_{-0.0497}$ & 0.8083 & $0.7774^{+0.0349}_{-0.0497}$\\
${\rm{Age}}/{\rm{Gyr}}$ & 13.46 & $13.59^{+0.0708}_{-0.143}$ & 13.78 & $13.79^{+0.0305}_{-0.0303}$ & 13.79 & $13.79^{+0.0352}_{-0.0351}$ & 13.78 & $13.79^{+0.0352}_{-0.0351}$
\end{tabular}
}
\bigskip
\resizebox{\textwidth}{!}{
\begin{tabular}{ccccc}
\hline
$\chi^2_{min} =$    & $\chi^2_{CMB} + \chi^2_{prior}$ & $\chi^2_{CMB} + \chi^2_{prior} + \chi^2_{BAO}$ & $\chi^2_{CMB} + \chi^2_{prior} + \chi^2_{SNIa}$ & $\chi^2_{CMB} + \chi^2_{prior} + \chi^2_{BAO} + \chi^2_{SNIa} + \chi^2_{H_0}$ \\
\cline{2-5}
     & 12930.22 + 11.92 &  12933.24 + 9.50 + 5.61 & 12933.39 + 11.00 + 696.05 & 12935.42 + 9.42 + 5.10 + 696.62 + 0.33 \\
\hline
\end{tabular}
}
\end{table*}
In Table \ref{tab.model2} we list the best fit and 68\% C.L. values for the parameters of Model II. The 1-D and 2-D marginalized posterior distributions are plotted in Figs. \ref{1d_dist2} and \ref{2d_dist2}. Similar to Model I, we do not observe any significant difference between the fitting results from Planck 2015 and the previous results obtained from  Planck 2013 alone. The joint analyses present some differences between the current results and the old ones reported in \cite{Costa:2013sva}. The main difference is that $\omega = -1$ can only be excluded in the $1\sigma$ range instead of $2\sigma$ as that in \cite{Costa:2013sva}. On the other hand, there is more room for the interaction to be allowed in the model with the most recent data sets.

Although we learnt that there are differences in observations between Planck 2013 and Planck 2015, especially at large scales, these differences are not strong enough to improve the constraints on the interaction models I and II. The reason behind is that theoretically at large scales there exists degeneracy between the coupling constant and the equation of state of dark energy in these two models \cite{He:2011qn,He:2009pd}. Near the first acoustic peak, the coupling constant is again degenerate with the dark matter abundance. Small differences between Planck 2015 and 2013 data at low $l$ are not effective enough to break these degeneracies so that the model parameters for Models I and II cannot be constrained much better. Some improvements in the model parameters discussed above can be attributed to the accuracy of the new data.

\begin{table*}[h!]
\caption{Cosmological parameters - Model III.}\centering\label{tab.model3}
\resizebox{\textwidth}{!}{
\begin{tabular}{lllllllll}
\hline
    & \multicolumn{2}{c}{Planck} & \multicolumn{2}{c}{Planck+BAO} & \multicolumn{2}{c}{Planck+SNIa} & \multicolumn{2}{c}{Planck+BAO+SNIa+H0} \\
    \cline{2-9}
    Parameter & Best fit  & 68\% limits & Best fit  & 68\% limits & Best fit  & 68\% limits & Best fit  & 68\% limits \\
    \hline
    $\Omega_b h^2$ & $0.0223$ & $0.02235^{+0.00017}_{-0.00017}$ & $0.02214$ & $0.0223^{+0.000164}_{-0.000178}$ & $0.02211$ & $0.02232^{+0.000173}_{-0.000174}$ & $0.02232$ & $0.02228^{+0.000154}_{-0.000175}$ \\
    $\Omega_c h^2$ & $0.1198$ & $0.1236^{+0.00235}_{-0.00353}$ & $0.1219$ & $0.1223^{+0.0014}_{-0.00165}$ & $0.1235$ & $0.1253^{+0.0027}_{-0.00364}$ & $0.121$ & $0.1216^{+0.00119}_{-0.00119}$ \\
    $100\theta_{MC}$ & $1.041$ & $1.041^{+0.000377}_{-0.000374}$ & $1.04$ & $1.041^{+0.000316}_{-0.000312}$ & $1.04$ & $1.04^{+0.000373}_{-0.000368}$ & $1.041$ & $1.041^{+0.000291}_{-0.000301}$ \\
    $\tau$ & $0.07784$ & $0.07051^{+0.0182}_{-0.0179}$ & $0.08062$ & $0.07547^{+0.0168}_{-0.0164}$ & $0.07695$ & $0.0711^{+0.0172}_{-0.017}$ & $0.06406$ & $0.07728^{+0.0169}_{-0.0167}$ \\
    $\ln (10^{10}A_s)$ & $3.087$ & $3.074^{+0.0357}_{-0.0355}$ & $3.097$ & $3.084^{+0.0326}_{-0.032}$ & $3.09$ & $3.075^{+0.0336}_{-0.0332}$ & $3.062$ & $3.088^{+0.0331}_{-0.0327}$ \\
    $n_s$ & $0.9649$ & $0.9608^{+0.00508}_{-0.00503}$ & $0.9614$ & $0.9617^{+0.00416}_{-0.00411}$ & $0.9604$ & $0.9589^{+0.00492}_{-0.00489}$ & $0.9618$ & $0.9624^{+0.00409}_{-0.00411}$ \\
    $w$ & $-1.701$ & $-1.702^{+0.298}_{-0.364}$ & $-1.139$ & $-1.167^{+0.0986}_{-0.0479}$ & $-1.078$ & $-1.132^{+0.0696}_{-0.0377}$ & $-1.06$ & $-1.104^{+0.0467}_{-0.0292}$ \\
    $\lambda_1$ & $0.0004372$ & $0.001458^{+0.000373}_{-0.00146}$ & $0.0007481$ & $0.0009446^{+0.000392}_{-0.000753}$ & $0.001086$ & $0.001828^{+0.000855}_{-0.00126}$ & $0.0007273$ & $0.0007127^{+0.000256}_{-0.000633}$ \\
    \hline
    $H_0$ & $89.51$ & $84.91^{+15.1}_{-4.8}$ & $69.56$ & $70.11^{+1.21}_{-1.86}$ & $66.85$ & $66.91^{+2.04}_{-1.86}$ & $67.93$ & $68.91^{+0.875}_{-0.997}$ \\
    $\Omega_{de}$ & $0.8218$ & $0.788^{+0.0686}_{-0.0268}$ & $0.701$ & $0.704^{+0.0116}_{-0.0138}$ & $0.6728$ & $0.6678^{+0.0292}_{-0.0214}$ & $0.688$ & $0.6955^{+0.009}_{-0.00886}$ \\
    $\Omega_m$ & $0.1782$ & $0.212^{+0.0268}_{-0.0686}$ & $0.299$ & $0.296^{+0.0138}_{-0.0116}$ & $0.3272$ & $0.3322^{+0.0214}_{-0.0292}$ & $0.312$ & $0.3045^{+0.00886}_{-0.009}$ \\
    $\sigma_8$ & $1.016$ & $0.9885^{+0.102}_{-0.061}$ & $0.8666$ & $0.8654^{+0.0181}_{-0.023}$ & $0.8464$ & $0.8432^{+0.0184}_{-0.0183}$ & $0.8293$ & $0.8535^{+0.0156}_{-0.0169}$ \\
    Age/Gyr & $13.55$ & $13.71^{+0.102}_{-0.18}$ & $13.84$ & $13.83^{+0.0417}_{-0.0462}$ & $13.91$ & $13.95^{+0.0908}_{-0.122}$ & $13.85$ & $13.83^{+0.0353}_{-0.0434}$
\end{tabular}
}
\bigskip
\resizebox{\textwidth}{!}{
\begin{tabular}{ccccc}
\hline
$\chi^2_{min} =$    & $\chi^2_{CMB} + \chi^2_{prior}$ & $\chi^2_{CMB} + \chi^2_{prior} + \chi^2_{BAO}$ & $\chi^2_{CMB} + \chi^2_{prior} + \chi^2_{SNIa}$ & $\chi^2_{CMB} + \chi^2_{prior} + \chi^2_{BAO} + \chi^2_{SNIa} + \chi^2_{H_0}$ \\
\cline{2-5}
     & 12933.78 + 9.74 &  12937.59 + 8.33 + 6.80 & 12935.62 + 12.68 + 695.84 & 12936.81 + 10.18 + 5.95 + 696.93 + 0.68 \\
\hline
\end{tabular}
}
\end{table*}
Now we move on to discuss the constraints on Model III, which is presented in Table \ref{tab.model3}. The 1-D and 2-D posterior distributions are plotted in Figs. \ref{1d_dist3} and \ref{2d_dist3}, respectively. It is clear that the fittings with Planck 2015 data alone present us a much better constraint on the model parameters as shown in Fig. \ref{1d_dist3} if we compare with the previous analysis by using Planck 2013 results. This is quite obvious in the $1\sigma$ range. If we add low redshift measurements, the new constraints for $\omega$ have become much closer to $-1$, but $\omega = -1$ is still excluded in more than 99\% C.L.. The interaction has been constrained to a smaller positive value with the new data sets compared with the previous tests.

\begin{table*}[h!]
\caption{Cosmological parameters - Model IV.}\centering\label{tab.model4}
\resizebox{\textwidth}{!}{
\begin{tabular}{lllllllll}
\hline
    & \multicolumn{2}{c}{Planck} & \multicolumn{2}{c}{Planck+BAO} & \multicolumn{2}{c}{Planck+SNIa} & \multicolumn{2}{c}{Planck+BAO+SNIa+H0} \\
    \cline{2-9}
    Parameter & Best fit  & 68\% limits & Best fit  & 68\% limits & Best fit  & 68\% limits & Best fit  & 68\% limits \\
    \hline
    $\Omega_b h^2$ & $0.0223$ & $0.02235^{+0.000178}_{-0.000179}$ & $0.02237$ & $0.0223^{+0.000167}_{-0.000167}$ & $0.02234$ & $0.02233^{+0.000172}_{-0.000175}$ & $0.02235$ & $0.02228^{+0.000161}_{-0.00016}$ \\
    $\Omega_c h^2$ & $0.1209$ & $0.124^{+0.0025}_{-0.0039}$ & $0.1216$ & $0.1226^{+0.00138}_{-0.00176}$ & $0.1236$ & $0.1255^{+0.00287}_{-0.00362}$ & $0.1212$ & $0.1218^{+0.00125}_{-0.00133}$ \\
    $100\theta_{MC}$ & $1.041$ & $1.041^{+0.000375}_{-0.000373}$ & $1.041$ & $1.041^{+0.000317}_{-0.000319}$ & $1.041$ & $1.04^{+0.000371}_{-0.000377}$ & $1.041$ & $1.041^{+0.000305}_{-0.000325}$ \\
    $\tau$ & $0.084$ & $0.07043^{+0.018}_{-0.0176}$ & $0.07245$ & $0.07552^{+0.0171}_{-0.0172}$ & $0.09833$ & $0.07173^{+0.0172}_{-0.017}$ & $0.09174$ & $0.07709^{+0.0166}_{-0.0165}$ \\
    $\ln (10^{10}A_s)$ & $3.1$ & $3.073^{+0.0351}_{-0.0344}$ & $3.073$ & $3.084^{+0.0338}_{-0.0338}$ & $3.129$ & $3.076^{+0.0335}_{-0.0335}$ & $3.121$ & $3.087^{+0.0321}_{-0.0321}$ \\
    $n_s$ & $0.9634$ & $0.9609^{+0.00512}_{-0.00518}$ & $0.9661$ & $0.9619^{+0.00428}_{-0.00426}$ & $0.9584$ & $0.9592^{+0.00481}_{-0.00479}$ & $0.9645$ & $0.9624^{+0.00417}_{-0.00416}$ \\
    $w$ & $-1.674$ & $-1.691^{+0.318}_{-0.359}$ & $-1.182$ & $-1.165^{+0.0955}_{-0.0473}$ & $-1.081$ & $-1.132^{+0.0682}_{-0.0379}$ & $-1.067$ & $-1.105^{+0.0468}_{-0.0288}$ \\
    $\lambda$ & $0.0007646$ & $0.001416^{+0.000365}_{-0.00142}$ & $0.001101$ & $0.0009529^{+0.000429}_{-0.000726}$ & $0.001234$ & $0.001807^{+0.00084}_{-0.00121}$ & $0.0002468$ & $0.000735^{+0.000254}_{-0.000679}$ \\
    \hline
    $H_0$ & $87.25$ & $84.63^{+15.4}_{-4.9}$ & $71.03$ & $70.02^{+1.24}_{-1.79}$ & $67.08$ & $66.97^{+1.97}_{-1.75}$ & $68.66$ & $68.88^{+0.854}_{-0.97}$ \\
    $\Omega_{de}$ & $0.8111$ & $0.7859^{+0.07}_{-0.0275}$ & $0.7133$ & $0.7028^{+0.012}_{-0.0132}$ & $0.6743$ & $0.6678^{+0.0282}_{-0.0209}$ & $0.6941$ & $0.6947^{+0.0088}_{-0.0089}$ \\
    $\Omega_m$ & $0.1889$ & $0.2141^{+0.0275}_{-0.07}$ & $0.2867$ & $0.2972^{+0.0132}_{-0.012}$ & $0.3257$ & $0.3322^{+0.0209}_{-0.0282}$ & $0.3059$ & $0.3053^{+0.0089}_{-0.0088}$ \\
    $\sigma_8$ & $1.006$ & $0.9833^{+0.102}_{-0.0636}$ & $0.8586$ & $0.8631^{+0.0193}_{-0.0228}$ & $0.8567$ & $0.8411^{+0.0183}_{-0.0187}$ & $0.8657$ & $0.852^{+0.0157}_{-0.0158}$ \\
    Age/Gyr & $13.61$ & $13.71^{+0.102}_{-0.176}$ & $13.81$ & $13.83^{+0.0417}_{-0.0474}$ & $13.9$ & $13.94^{+0.0875}_{-0.119}$ & $13.79$ & $13.83^{+0.0354}_{-0.0439}$
\end{tabular}
}
\bigskip
\resizebox{\textwidth}{!}{
\begin{tabular}{ccccc}
\hline
$\chi^2_{min} =$    & $\chi^2_{CMB} + \chi^2_{prior}$ & $\chi^2_{CMB} + \chi^2_{prior} + \chi^2_{BAO}$ & $\chi^2_{CMB} + \chi^2_{prior} + \chi^2_{SNIa}$ & $\chi^2_{CMB} + \chi^2_{prior} + \chi^2_{BAO} + \chi^2_{SNIa} + \chi^2_{H_0}$ \\
\cline{2-5}
     & 12931.37 + 11.90 &  12936.13 + 7.80 + 8.33 & 12936.11 + 10.90 + 695.65 & 12937.13 + 10.61 + 5.53 + 695.74 + 0.37 \\
\hline
\end{tabular}
}
\end{table*}
Now we report the fitting results for Model IV with the new data sets. In Table \ref{tab.model4} we list detailed information on the constraints of the model parameters. The most significant 1-D and 2-D posterior distributions are plotted in Figs. \ref{1d_dist4} and \ref{2d_dist4}, respectively.  The improvements on the constraints of the model parameters are similar to Model III. The major difference we observe is that the new data sets allow more room for a nonzero interaction than that obtained by using the old data \cite{Costa:2013sva}. In the analysis using Planck 2013 data, the constrained coupling strength for Model III and Model IV are distinct: the posterior peak of $\lambda_1$ in Model III is at $\sim 10^{-3}$, while in Model IV it strongly prefers $\lambda \sim 0$. However, the results in the analysis using Planck 2015 data for these two models become similar. In both models the coupling coefficient peaks appear at $\sim 10^{-3}$ though still the null interactions are not ruled out completely.

For Models III and IV, the physics presented in \cite{He:2010im} told us that for these two models the coupling constant can be distinguished from the dark energy equation of state at small $l$ CMB spectrum. This is why we have tighter constraints on the coupling constant in these two models than in Models I and II. Besides, the tight constraint on the coupling constant can also help in turn to break the degeneracy in model parameters and get better constraints on dark energy equation of state and dark matter abundance. With the better quality of the Planck 2015 data sets and their differences from Planck 2013 result, especially at small $l$, the model parameters in Models III and IV can be better constrained compared with the old results.

In the following discussion, we add new complementary data sets from large scale structure observations, the redshift-space distortions data in Table \ref{RSD}, to investigate the constraints on interacting dark energy models from the joint analysis of BAO + SNIa + $H_0$ + RSD\footnote{In this case, we fix the parameters $\tau$, $A_s$ and $n_s$ to the mean value from Planck 2015.}, Planck 2015 + RSD and Planck 2015 + BAO + SNIa + $H_0$ + RSD. The behaviors of the four interacting models confronting to RSD data alone produce results similar to those by employing other low redshift measurements, but with broader uncertainties due to the big error bars in the RSD data observations. Thus it is still premature to use the RSD data alone in constraining the cosmological models. But this does not reduce the importance to include the RSD data as a complementary joint test together with other observational data to examine the cosmological models. In Fig. \ref{fsigma8}, we observe that when the interaction is large enough, it can lead to an accelerated growth of structures or even a decrease of the growth at late times for different interacting models. Thus the RSD data are useful to be used to effectively test the interacting dark energy models.

\begin{table*}[h!]
\caption{Cosmological parameters using RSD data - Model I.}\centering\label{tab.model1_RSD}
\resizebox{\textwidth}{!}{
\begin{tabular}{lllllllll}
\hline
    & \multicolumn{2}{c}{Planck} & \multicolumn{2}{c}{BAO+SNIa+H0+RSD} & \multicolumn{2}{c}{Planck+RSD} & \multicolumn{2}{c}{Planck+BAO+SNIa+H0+RSD} \\
    \cline{2-9}
    Parameter & Best fit & 68\% limits & Best fit & 68\% limits & Best fit & 68\% limits & Best fit & 68\% limits\\
    \hline
    $\Omega_b h^2$ & $0.02231$ & $0.0222^{+0.00016}_{-0.00016}$& $0.03126$ & $0.02173^{+0.00497}_{-0.0167}$& $0.02216$ & $0.02224^{+0.000158}_{-0.000159}$& $0.02233$ & $0.02237^{+0.000142}_{-0.000142}$ \\
    $\Omega_c h^2$ & $0.04788$ & $0.07131^{+0.0472}_{-0.024}$& $0.11$ & $0.103^{+0.00708}_{-0.0056}$& $0.1194$ & $0.1185^{+0.00162}_{-0.0016}$& $0.1172$ & $0.117^{+0.00134}_{-0.00123}$ \\
    $100\theta_{MC}$ & $1.045$ & $1.044^{+0.0015}_{-0.00329}$& $1.013$ & $1.035^{+0.03}_{-0.0191}$& $1.041$ & $1.041^{+0.000319}_{-0.000343}$& $1.041$ & $1.041^{+0.000306}_{-0.000308}$ \\
    $\tau$ & $0.08204$ & $0.08063^{+0.0171}_{-0.0169}$& - & - & $0.07073$ & $0.07198^{+0.0177}_{-0.0171}$& $0.05918$ & $0.06583^{+0.016}_{-0.0161}$ \\
    $\ln (10^{10}A_s)$ & $3.102$ & $3.097^{+0.0328}_{-0.0329}$& - & - & $3.074$ & $3.077^{+0.0337}_{-0.033}$& $3.044$ & $3.061^{+0.0309}_{-0.0311}$ \\
    $n_s$ & $0.9639$ & $0.9633^{+0.00472}_{-0.00514}$& - & - & $0.9649$ & $0.9648^{+0.00481}_{-0.00477}$& $0.9718$ & $0.9683^{+0.00431}_{-0.00422}$ \\
    $w$ & $-0.9765$ & $-0.9031^{+0.023}_{-0.0959}$& $-0.9891$ & $-0.9022^{+0.0298}_{-0.0968}$& $-0.8446$ & $-0.7841^{+0.0809}_{-0.083}$& $-0.975$ & $-0.9541^{+0.0188}_{-0.0372}$ \\
    $\lambda_2$ & $-0.1831$ & $-0.1297^{+0.13}_{-0.0448}$& $-0.002222$ & $-0.006925^{+0.00692}_{-0.00171}$& $-0.002074$ & $-0.002566^{+0.00257}_{-0.000426}$& $-0.0009821$ & $-0.001815^{+0.00182}_{-0.000328}$ \\
    \hline
    $H_0$ & $72.36$ & $68.1^{+3.99}_{-3.2}$& $68.15$ & $67.84^{+0.972}_{-0.973}$& $62.7$ & $61.32^{+2.48}_{-2.43}$& $67.47$ & $66.85^{+0.861}_{-0.713}$ \\
    $\Omega_{de}$ & $0.8647$ & $0.7899^{+0.0932}_{-0.106}$& $0.6944$ & $0.7274^{+0.0362}_{-0.0302}$& $0.6383$ & $0.6221^{+0.0356}_{-0.0294}$& $0.692$ & $0.6865^{+0.0087}_{-0.00755}$ \\
    $\Omega_m$ & $0.1353$ & $0.2101^{+0.106}_{-0.0926}$& $0.3056$ & $0.2726^{+0.0302}_{-0.0362}$& $0.3617$ & $0.3779^{+0.0294}_{-0.0356}$& $0.308$ & $0.3135^{+0.00755}_{-0.0087}$ \\
    $\sigma_8$ & $1.622$ & $1.438^{+0.143}_{-0.789}$& $0.739$ & $0.7633^{+0.0398}_{-0.039}$& $0.7817$ & $0.7622^{+0.0216}_{-0.0215}$& $0.7996$ & $0.8003^{+0.0123}_{-0.0123}$ \\
    Age/Gyr & $13.71$ & $13.81^{+0.058}_{-0.0916}$& $13.72$ & $13.97^{+0.268}_{-0.268}$& $13.93$ & $13.97^{+0.0717}_{-0.089}$& $13.8$ & $13.82^{+0.0235}_{-0.0235}$
\end{tabular}
}
\bigskip
\resizebox{\textwidth}{!}{
\begin{tabular}{ccccc}
\hline
$\chi^2_{min} =$    & $\chi^2_{CMB} + \chi^2_{prior}$ & $\chi^2_{BAO} + \chi^2_{SNIa} + \chi^2_{H_0} + \chi^2_{RSD}$ & $\chi^2_{CMB} + \chi^2_{prior} + \chi^2_{RSD}$ & $\chi^2_{CMB} + \chi^2_{prior} + \chi^2_{BAO} + \chi^2_{SNIa} + \chi^2_{H_0} + \chi^2_{RSD}$ \\
\cline{2-5}
     & 12935.29 + 10.89 & 11.84 + 695.28 + 0.54 + 7.51 & 12935.90 + 11.04 + 16.11 & 12941.97 + 11.24 + 18.48 + 695.66 + 0.87 + 14.43 \\
\hline
\end{tabular}
}
\end{table*}

\begin{figure}[h!]
\subfloat[Model I]{
\includegraphics[width=\textwidth]{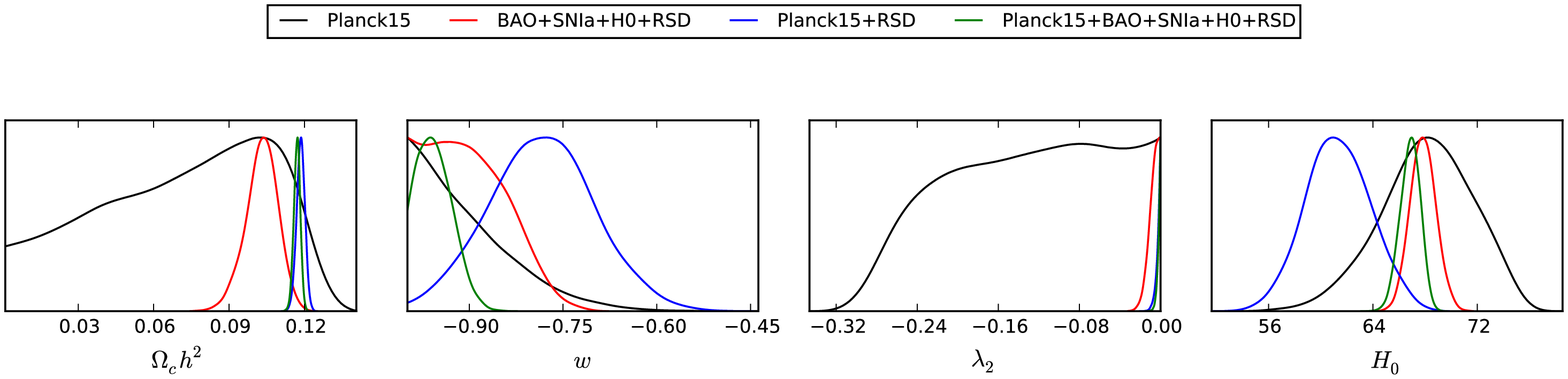}\label{1d_dist1_RSD}}\\
\subfloat[Model II]{
\includegraphics[width=\textwidth]{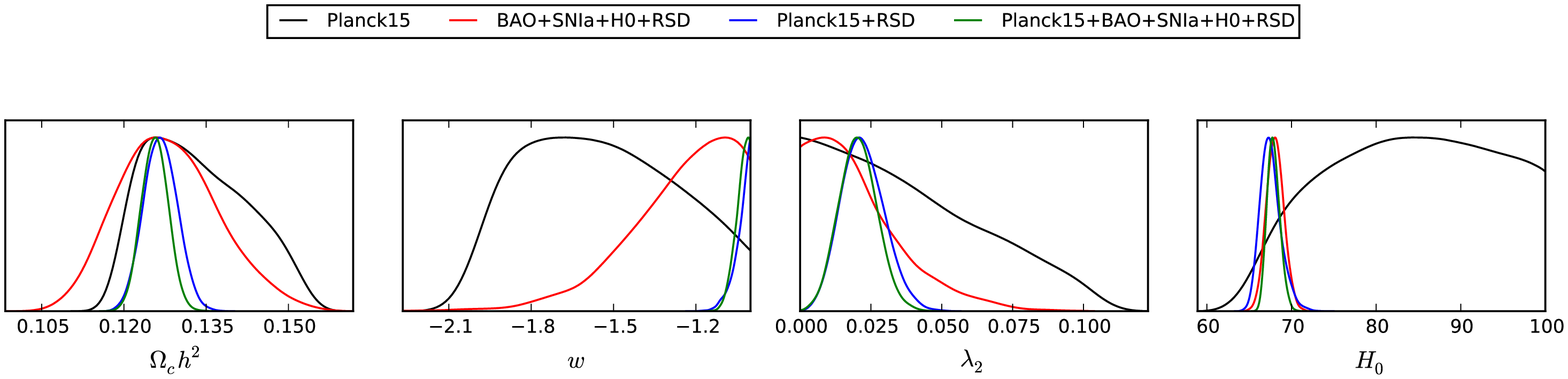}\label{1d_dist2_RSD}}\\
\subfloat[Model III]{
\includegraphics[width=\textwidth]{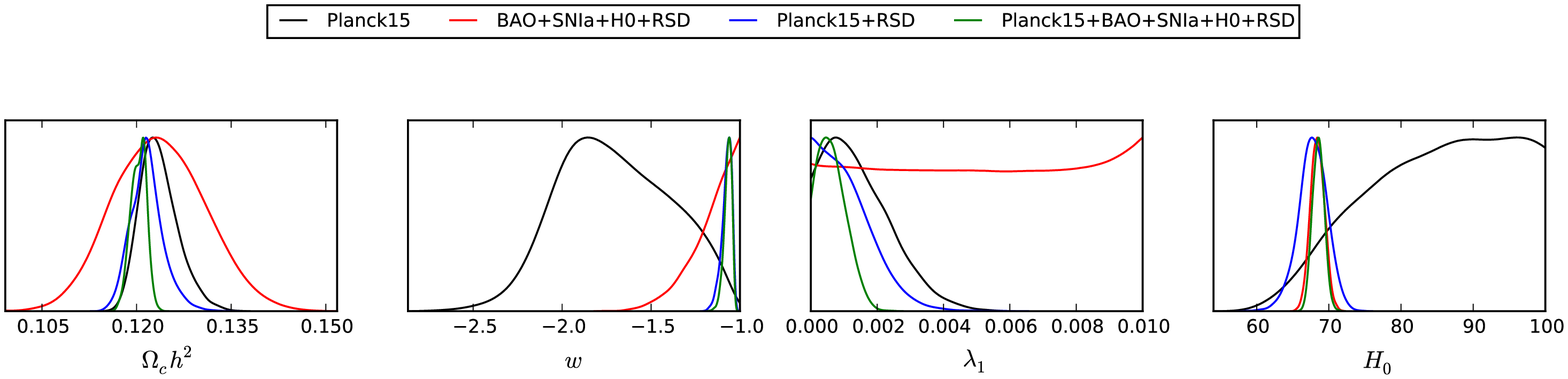}\label{1d_dist3_RSD}}\\
\subfloat[Model IV]{
\includegraphics[width=\textwidth]{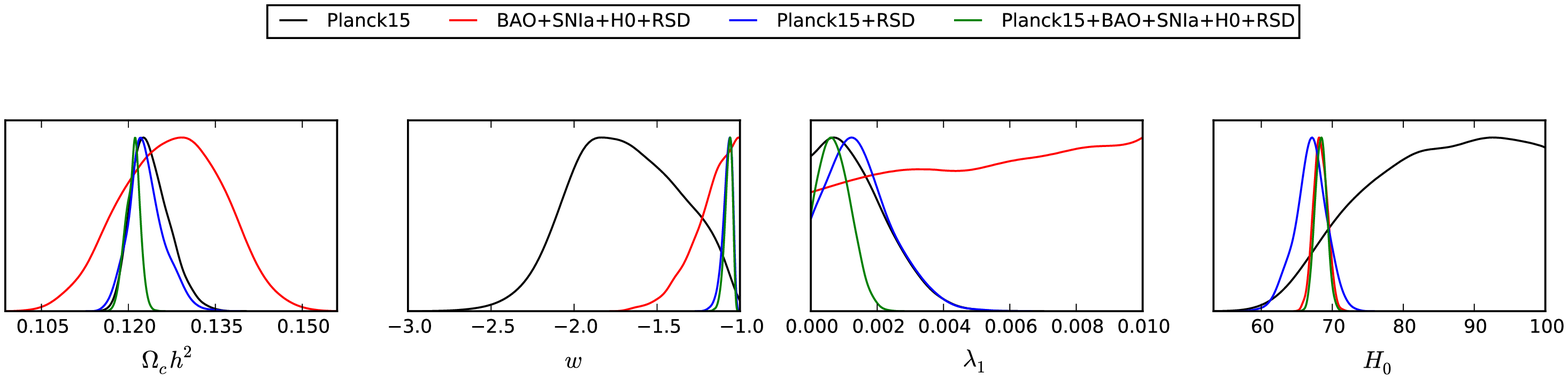}\label{1d_dist4_RSD}}
\caption{1-D distribution for selected parameters using RSD data.}
\end{figure}

\begin{figure}[h!]
\subfloat[Model I]{
\includegraphics[width=0.85\textwidth]{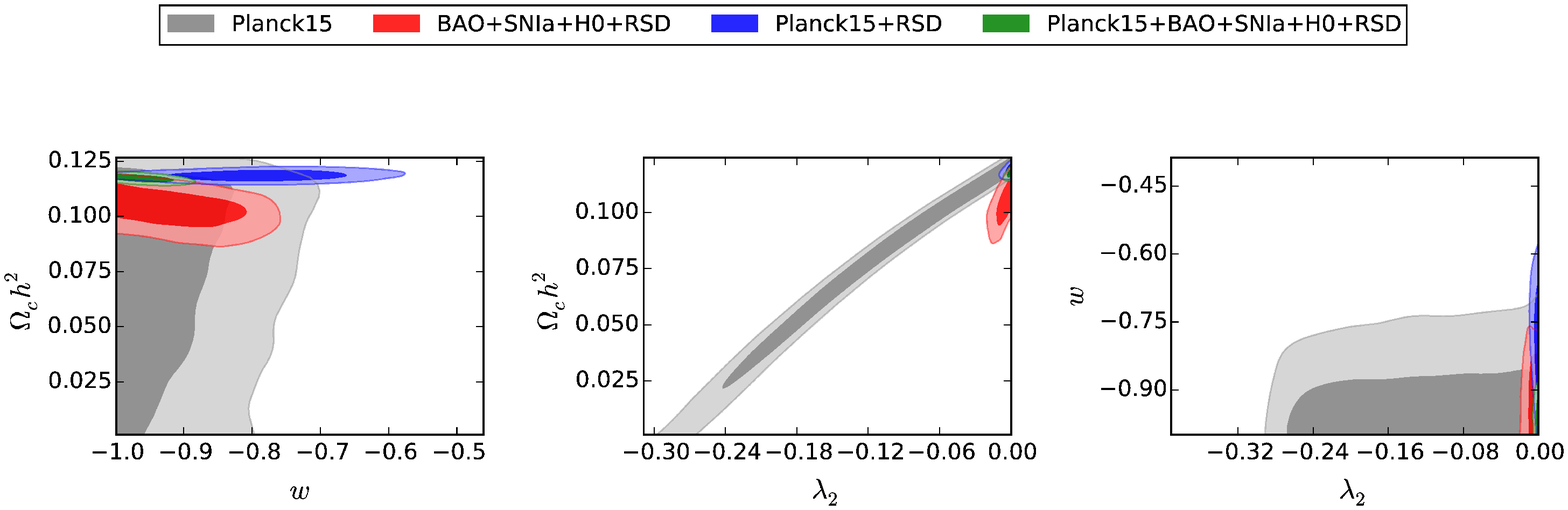}\label{2d_dist1_RSD}}\\
\subfloat[Model II]{
\includegraphics[width=0.85\textwidth]{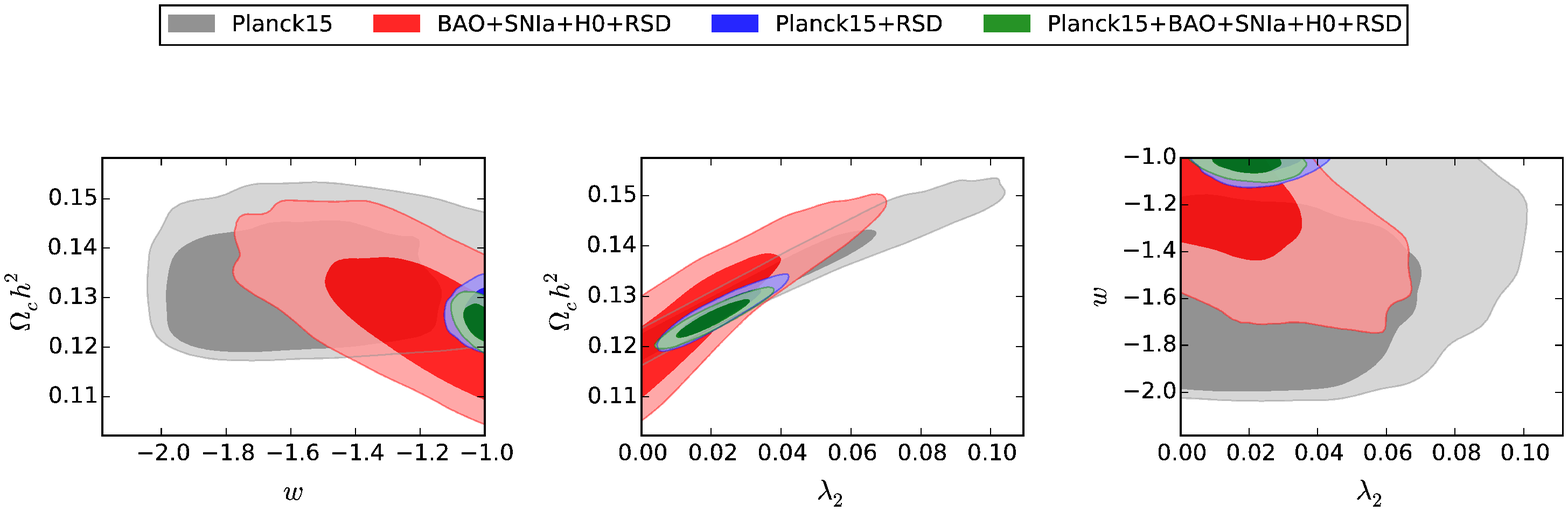}\label{2d_dist2_RSD}}\\
\subfloat[Model III]{
\includegraphics[width=0.85\textwidth]{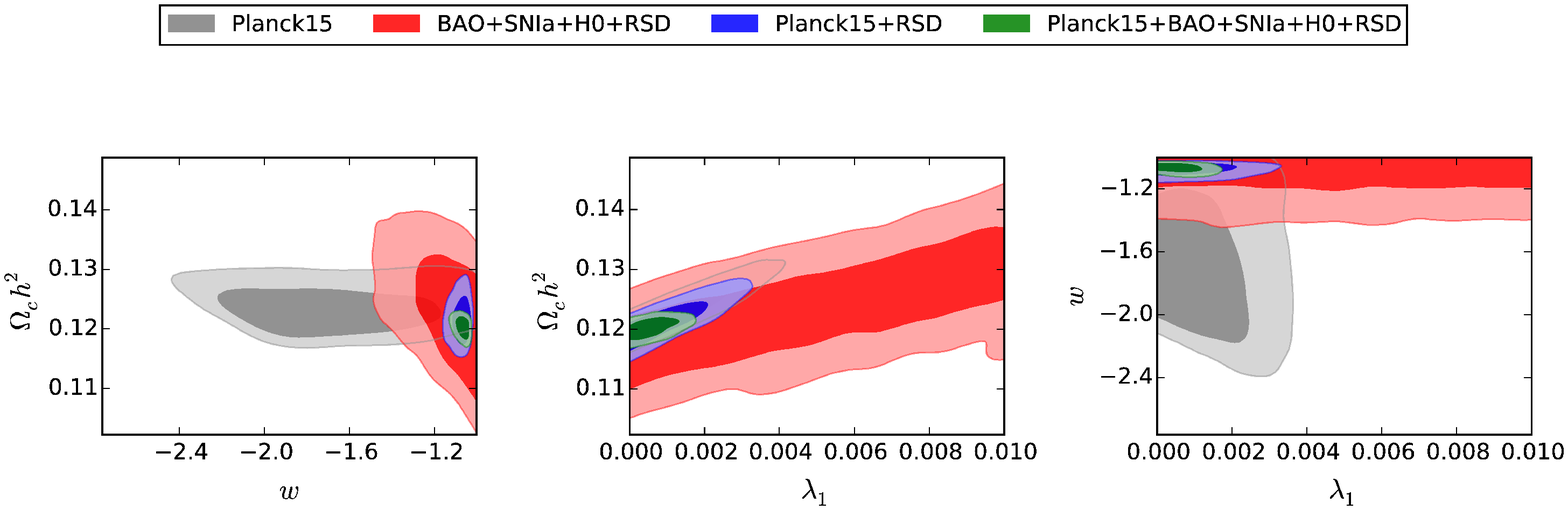}\label{2d_dist3_RSD}}\\
\subfloat[Model IV]{
\includegraphics[width=0.85\textwidth]{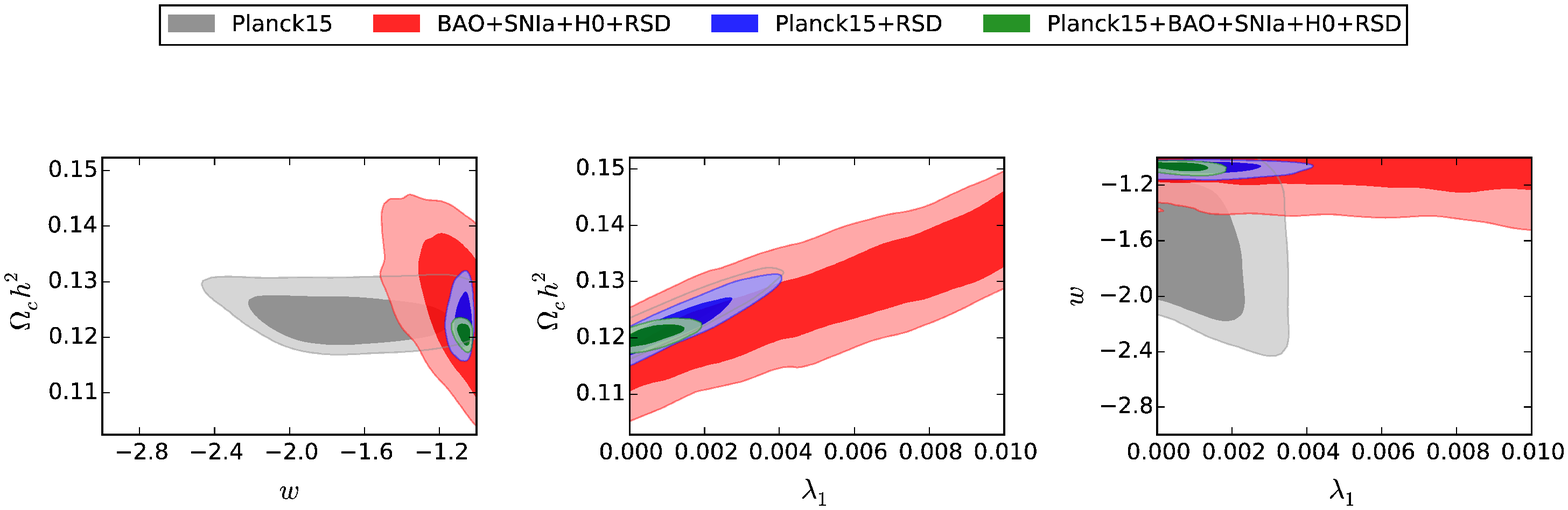}\label{2d_dist4_RSD}}
\caption{2-D distribution for selected parameters using RSD data.}
\end{figure}
\begin{figure}[h!]
\includegraphics[width=\textwidth]{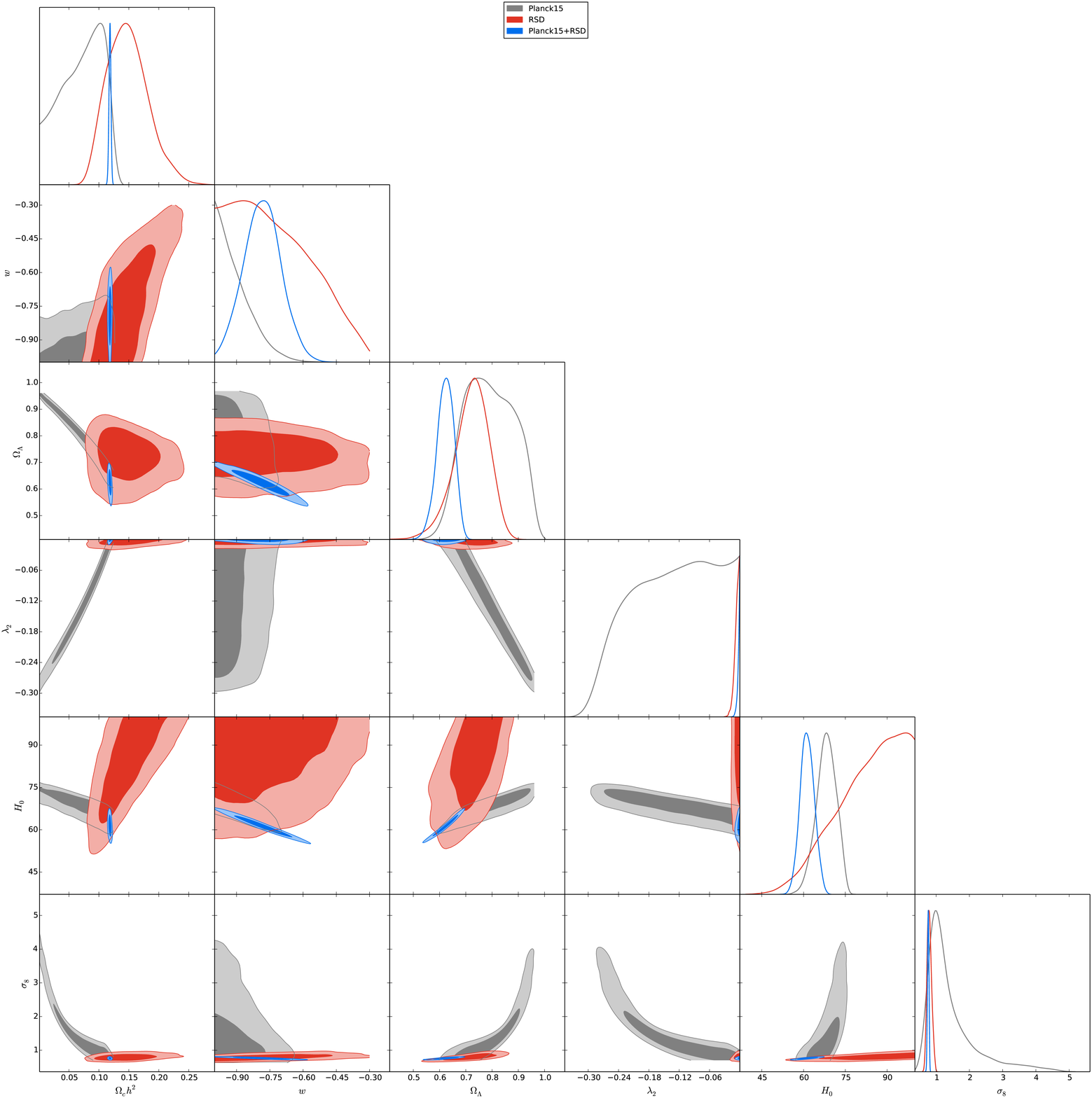}
\caption{1-D and 2-D distributions for Model I with Planck and RSD data.}
\label{Model1_tri}
\end{figure}
\begin{figure}[h!]
\subfloat[Planck15]{
\includegraphics[width=0.5\textwidth]{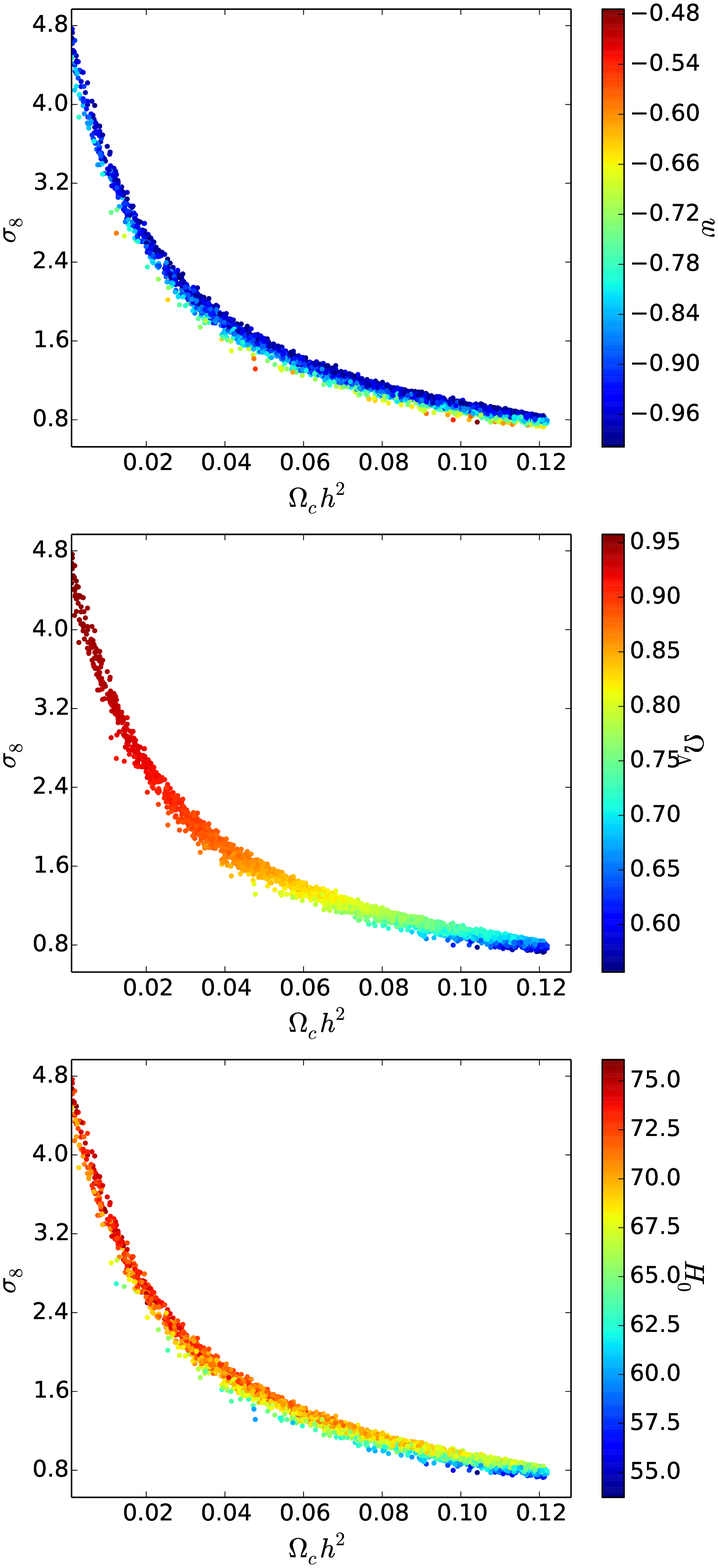}\label{Model1_P_3D}}
\subfloat[Planck15+RSD]{
\includegraphics[width=0.5\textwidth]{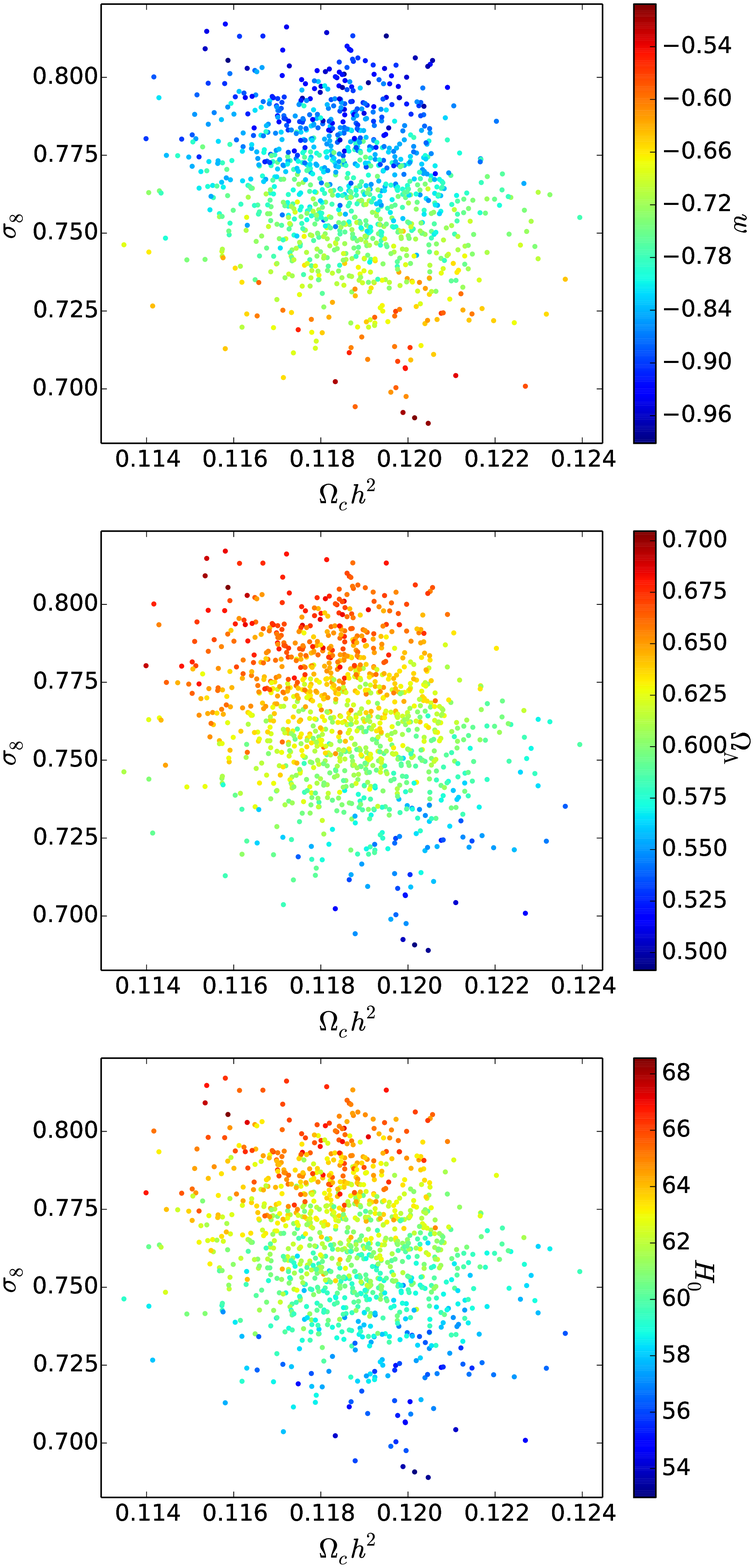}\label{Model1_PR_3D}}
\caption{3-D distribution for selected parameters of Model I.}
\end{figure}

In Table \ref{tab.model1_RSD} we present the best fit and 68\% C.L. limits by combining RSD data to constrain Model I. The 1-D and 2-D posterior distributions are plotted in Figs. \ref{1d_dist1_RSD} and \ref{2d_dist1_RSD}, respectively. For better comparison, we also include the result by using Planck 2015 data again. We see that the constraints from the low redshift measurements are consistent with the CMB measurements but with a narrower posterior for the parameters than the CMB observation. In the discussion above we found that the Planck result adding other measurements can put a preference for a negative interaction for Model I, while including the RSD data we see that the interaction has been excluded with high probability. This effect can be attributed to the fact that the growth factor for Model I can grow at late times, which is actually not observed from RSD data. Similar result was also obtained in \cite{Yang:2014gza}. In \cite{Salvatelli:2014zta}, a special interaction model with a vacuum energy interacting with dark matter was investigated and it was found that when the interaction is time-dependent and proportional to the energy density of dark energy, the nonzero interaction between dark sectors is still allowed by the RSD data. The joint analysis by combing the RSD data together with the Planck data presents us a strange result, the nonzero interaction is no longer allowed but the mean value of dark energy equation of state tends to $\omega = -0.7841$ in conflict with the $\Lambda$CDM result in more than 99\% C.L. The inclusion of other measurements can alleviate this tension by allowing the dark energy equation of state closer to $\omega = -1$. To better understand this effect, we plot in Fig. \ref{Model1_tri} some distributions for RSD alone\footnote{Here we also fix $\tau$, $A_s$ and $n_s$ to the mean value from Planck 2015.}, together with the distributions from Planck 2015 and Planck + RSD. What we can see is that there is $\sim 1\sigma$ tension between the values for $\Omega_ch^2$ from RSD and Planck alone, the RSD data preferring higher values. Thus, in the joint analysis, the value of $\Omega_ch^2$ is tight constrained to $\Omega_ch^2 \approx 0.12$. On the other hand, the value of $\sigma_8$ is tight constrained to $\sigma_8 \approx 0.76$. Thus, looking at Fig. \ref{Model1_P_3D}, we can see that those values correspond to the right end of the $\sigma_8$ x $\Omega_ch^2$ curves, where there is less room for $\omega$ close to -1. Figure \ref{Model1_PR_3D} shows those curves for the Planck + RSD joint analysis, which can be thought as a zoom in Fig. \ref{Model1_P_3D}.

\begin{table*}[h!]
\caption{Cosmological parameters using RSD data - Model II.}\centering\label{tab.model2_RSD}
\resizebox{\textwidth}{!}{
\begin{tabular}{lllllllll}
\hline
    & \multicolumn{2}{c}{Planck} & \multicolumn{2}{c}{BAO+SNIa+H0+RSD} & \multicolumn{2}{c}{Planck+RSD} & \multicolumn{2}{c}{Planck+BAO+SNIa+H0+RSD} \\
    \cline{2-9}
    Parameter & Best fit & 68\% limits & Best fit & 68\% limits & Best fit & 68\% limits & Best fit & 68\% limits\\
    \hline
    $\Omega_b h^2$ & $0.02232$ & $0.02225^{+0.000162}_{-0.000161}$& $0.03769$ & $0.05248^{+0.0137}_{-0.0176}$& $0.02221$ & $0.0222^{+0.000157}_{-0.000156}$& $0.02243$ & $0.02226^{+0.000138}_{-0.000139}$ \\
    $\Omega_c h^2$ & $0.1314$ & $0.1334^{+0.00692}_{-0.0125}$& $0.1172$ & $0.1282^{+0.00808}_{-0.0103}$& $0.1245$ & $0.1267^{+0.00302}_{-0.00298}$& $0.1246$ & $0.1256^{+0.00238}_{-0.00243}$ \\
    $100\theta_{MC}$ & $1.04$ & $1.04^{+0.000651}_{-0.000562}$& $0.9973$ & $0.9656^{+0.0346}_{-0.0346}$& $1.041$ & $1.04^{+0.000355}_{-0.000353}$& $1.04$ & $1.04^{+0.000315}_{-0.000318}$ \\
    $\tau$ & $0.07543$ & $0.07653^{+0.0177}_{-0.0174}$& - & - & $0.0553$ & $0.06963^{+0.0165}_{-0.0166}$& $0.0715$ & $0.07244^{+0.0164}_{-0.0164}$ \\
    $\ln (10^{10}A_s)$ & $3.082$ & $3.088^{+0.0342}_{-0.0337}$& - & - & $3.044$ & $3.074^{+0.0317}_{-0.0321}$& $3.078$ & $3.078^{+0.0321}_{-0.0321}$ \\
    $n_s$ & $0.9657$ & $0.9638^{+0.00477}_{-0.00475}$& - & - & $0.9637$ & $0.9628^{+0.00472}_{-0.00479}$& $0.9684$ & $0.9647^{+0.00409}_{-0.00413}$ \\
    $w$ & $-1.872$ & $-1.55^{+0.235}_{-0.358}$& $-1.051$ & $-1.264^{+0.263}_{-0.063}$& $-1.05$ & $-1.036^{+0.035}_{-0.00703}$& $-1.005$ & $-1.035^{+0.0341}_{-0.00835}$ \\
    $\lambda_2$ & $0.02931$ & $0.03884^{+0.0116}_{-0.0388}$& $0.005695$ & $0.02085^{+0.00398}_{-0.0209}$& $0.01486$ & $0.02215^{+0.007}_{-0.0084}$& $0.01935$ & $0.02047^{+0.00656}_{-0.00667}$ \\
    \hline
    $H_0$ & $96.2$ & $83.88^{+13.3}_{-7.86}$& $67.91$ & $68.07^{+1.}_{-1.02}$& $68.27$ & $67.55^{+0.949}_{-1.43}$& $67.45$ & $67.89^{+0.661}_{-0.842}$ \\
    $\Omega_{de}$ & $0.8331$ & $0.7688^{+0.0778}_{-0.0353}$& $0.6628$ & $0.6084^{+0.0521}_{-0.0469}$& $0.6838$ & $0.6718^{+0.0162}_{-0.0163}$& $0.6755$ & $0.6775^{+0.0098}_{-0.01}$ \\
    $\Omega_m$ & $0.1669$ & $0.2312^{+0.0353}_{-0.0778}$& $0.3372$ & $0.3916^{+0.0469}_{-0.0521}$& $0.3162$ & $0.3282^{+0.0163}_{-0.0162}$& $0.3245$ & $0.3225^{+0.01}_{-0.0098}$ \\
    $\sigma_8$ & $0.9852$ & $0.9016^{+0.0945}_{-0.094}$& $0.7216$ & $0.6998^{+0.0344}_{-0.0435}$& $0.8004$ & $0.7977^{+0.0134}_{-0.0151}$& $0.7895$ & $0.7991^{+0.0129}_{-0.0128}$ \\
    Age/Gyr & $13.46$ & $13.59^{+0.0708}_{-0.143}$& $13.55$ & $13.24^{+0.301}_{-0.318}$& $13.79$ & $13.81^{+0.0372}_{-0.0315}$& $13.8$ & $13.8^{+0.0227}_{-0.0227}$
\end{tabular}
}
\bigskip
\resizebox{\textwidth}{!}{
\begin{tabular}{ccccc}
\hline
$\chi^2_{min} =$    & $\chi^2_{CMB} + \chi^2_{prior}$ & $\chi^2_{BAO} + \chi^2_{SNIa} + \chi^2_{H_0} + \chi^2_{RSD}$ & $\chi^2_{CMB} + \chi^2_{prior} + \chi^2_{RSD}$ & $\chi^2_{CMB} + \chi^2_{prior} + \chi^2_{BAO} + \chi^2_{SNIa} + \chi^2_{H_0} + \chi^2_{RSD}$ \\
\cline{2-5}
     & 12930.22 + 11.92 & 11.12 + 695.78 + 0.66 + 7.25 & 12935.95 + 9.74 + 12.07 & 12938.57 + 8.86 + 14.56 + 696.08 + 0.89 + 10.33 \\
\hline
\end{tabular}
}
\end{table*}
The constraints for Model II by employing the RSD data are reported in Table \ref{tab.model2_RSD} and the 1-D and 2-D posterior distributions are shown in Figs. \ref{1d_dist2_RSD} and \ref{2d_dist2_RSD}. We observe that the combination of RSD with Planck data shows a preference for the dark energy equation of state close to $-1$ but keeps the interaction parameter positive by excluding the null interaction possibility. The joint analysis including all data establishes the best fit value for the coupling $\lambda_2 = 0.01935$ with mean $\lambda_2 = 0.02047$, and excludes a null interaction in more than 99\% C.L.. In a recent paper \cite{Murgia:2016ccp}, the authors also obtained a result consistent with a positive interaction, although they did a different analysis for the RSD data. They used a combination of BAO/RSD data in addition to Planck 2015. The BAO data used in \cite{Murgia:2016ccp} are the same as the BAO data sets in this work and the RSD data were taken from \cite{Samushia:2013yga}. The positive interaction is interesting to us, since it can help alleviate the coincidence problem as there is longer time for the energy densities of dark matter and dark energy to be comparable. Therefore, Model II arises as a possible candidate to solve the coincidence problem in light of the new complementary RSD data.

\begin{table*}[h!]
\caption{Cosmological parameters using RSD data - Model III.}\centering\label{tab.model3_RSD}
\resizebox{\textwidth}{!}{
\begin{tabular}{lllllllll}
\hline
    & \multicolumn{2}{c}{Planck} & \multicolumn{2}{c}{BAO+SNIa+H0+RSD} & \multicolumn{2}{c}{Planck+RSD} & \multicolumn{2}{c}{Planck+BAO+SNIa+H0+RSD} \\
    \cline{2-9}
    Parameter & Best fit & 68\% limits & Best fit & 68\% limits & Best fit & 68\% limits & Best fit & 68\% limits\\
    \hline
    $\Omega_b h^2$ & $0.0223$ & $0.02235^{+0.00017}_{-0.00017}$ & $0.02324$ & $0.04002^{+0.0104}_{-0.0151}$ & $0.02225$ & $0.02233^{+0.000165}_{-0.000166}$ & $0.02235$ & $0.02232^{+0.000156}_{-0.000156}$  \\
    $\Omega_c h^2$ & $0.1198$ & $0.1236^{+0.00235}_{-0.00353}$ & $0.1268$ & $0.1236^{+0.00695}_{-0.00779}$ & $0.1196$ & $0.1217^{+0.00211}_{-0.00289}$ & $0.1194$ & $0.1203^{+0.00138}_{-0.00114}$  \\
    $100\theta_{MC}$ & $1.041$ & $1.041^{+0.000377}_{-0.000374}$ & $1.07$ & $1.009^{+0.0374}_{-0.0371}$ & $1.041$ & $1.041^{+0.000349}_{-0.000341}$ & $1.041$ & $1.041^{+0.000293}_{-0.00029}$  \\
    $\tau$ & $0.07784$ & $0.07051^{+0.0182}_{-0.0179}$ & - & - & $0.05039$ & $0.04261^{+0.0162}_{-0.0165}$ & $0.0483$ & $0.04598^{+0.0156}_{-0.0158}$  \\
    $\ln (10^{10}A_s)$ & $3.087$ & $3.074^{+0.0357}_{-0.0355}$ & - & - & $3.028$ & $3.014^{+0.0303}_{-0.0337}$ & $3.025$ & $3.021^{+0.03}_{-0.0307}$  \\
    $n_s$ & $0.9649$ & $0.9608^{+0.00508}_{-0.00503}$ & - & - & $0.962$ & $0.9622^{+0.0049}_{-0.00496}$ & $0.9652$ & $0.9634^{+0.00414}_{-0.00413}$  \\
    $w$ & $-1.701$ & $-1.702^{+0.298}_{-0.364}$ & $-1.064$ & $-1.159^{+0.158}_{-0.0354}$ & $-1.061$ & $-1.075^{+0.0362}_{-0.0173}$ & $-1.055$ & $-1.069^{+0.0268}_{-0.0152}$  \\
    $\lambda_1$ & $0.0004372$ & $0.001458^{+0.000373}_{-0.00146}$ & $0.009867$ & $0.005046^{+0.00495}_{-0.00505}$ & $0.0002835$ & $0.001104^{+0.000277}_{-0.0011}$ & $0.0003881$ & $0.0006628^{+0.000241}_{-0.000592}$  \\
    \hline
    $H_0$ & $89.51$ & $84.91^{+15.1}_{-4.8}$ & $68.34$ & $68.42^{+0.986}_{-0.957}$ & $68.86$ & $67.74^{+2.13}_{-1.83}$ & $68.87$ & $68.54^{+0.799}_{-0.824}$  \\
    $\Omega_{de}$ & $0.8218$ & $0.788^{+0.0686}_{-0.0268}$ & $0.6774$ & $0.6488^{+0.0416}_{-0.035}$ & $0.6995$ & $0.6837^{+0.0277}_{-0.0187}$ & $0.6998$ & $0.6948^{+0.00914}_{-0.00934}$  \\
    $\Omega_m$ & $0.1782$ & $0.212^{+0.0268}_{-0.0686}$ & $0.3226$ & $0.3512^{+0.035}_{-0.0416}$ & $0.3005$ & $0.3163^{+0.0187}_{-0.0277}$ & $0.3002$ & $0.3052^{+0.00934}_{-0.00913}$  \\
    $\sigma_8$ & $1.016$ & $0.9885^{+0.102}_{-0.061}$ & $0.731$ & $0.7032^{+0.0361}_{-0.0359}$ & $0.8193$ & $0.8091^{+0.0173}_{-0.0159}$ & $0.8155$ & $0.8146^{+0.0126}_{-0.0127}$  \\
    Age/Gyr & $13.55$ & $13.71^{+0.102}_{-0.18}$ & $13.72$ & $13.46^{+0.267}_{-0.271}$ & $13.81$ & $13.88^{+0.068}_{-0.111}$ & $13.8$ & $13.83^{+0.0352}_{-0.045}$
\end{tabular}
}
\bigskip
\resizebox{\textwidth}{!}{
\begin{tabular}{ccccc}
\hline
$\chi^2_{min} =$    & $\chi^2_{CMB} + \chi^2_{prior}$ & $\chi^2_{BAO} + \chi^2_{SNIa} + \chi^2_{H_0} + \chi^2_{RSD}$ & $\chi^2_{CMB} + \chi^2_{prior} + \chi^2_{RSD}$ & $\chi^2_{CMB} + \chi^2_{prior} + \chi^2_{BAO} + \chi^2_{SNIa} + \chi^2_{H_0} + \chi^2_{RSD}$ \\
\cline{2-5}
     & 12933.78 + 9.74 &  11.75 + 695.47 + 0.49 + 7.30 & 12944.86 + 9.23 + 21.98 & 12941.76 + 14.04 + 25.61 + 696.32 + 0.30 + 20.24 \\
\hline
\end{tabular}
}
\end{table*}
Unlike Models I and II, the interaction Models III and IV have not been tested with the RSD data in the available references. Here we are going to investigate the Models III and IV by confronting to RSD data sets and other external measurements. In Tables \ref{tab.model3_RSD} and \ref{tab.model4_RSD} we report the best fits and 68\% C.L. limits respectively. The 1-D and 2-D posterior distributions are shown in Figs. \ref{1d_dist3_RSD} and \ref{2d_dist3_RSD} for Model III, and Figs. \ref{1d_dist4_RSD} and \ref{2d_dist4_RSD} for Model IV. The results indicate that there is a tension between the expected dark energy equation of state from Planck and by using low redshift measurements alone. Planck data show a tendency for more negative values of $\omega$, while low redshift measurements give more consistent result with $\omega = -1$. Therefore, the joint analyses lead to an equation of state for dark energy close to minus one, but still exclude it in more than 99\% C.L. limit. There are rooms for small positive interaction in those models, although one order of magnitude smaller than the value obtained for Model II. Unlike Model II, the zero interaction cannot be clearly ruled out from the available data fitting.
\begin{table*}[h!]
\caption{Cosmological parameters using RSD data - Model IV.}\centering\label{tab.model4_RSD}
\resizebox{\textwidth}{!}{
\begin{tabular}{lllllllll}
\hline
    & \multicolumn{2}{c}{Planck} & \multicolumn{2}{c}{BAO+SNIa+H0+RSD} & \multicolumn{2}{c}{Planck+RSD} & \multicolumn{2}{c}{Planck+BAO+SNIa+H0+RSD} \\
    \cline{2-9}
    Parameter & Best fit & 68\% limits & Best fit & 68\% limits & Best fit & 68\% limits & Best fit & 68\% limits\\
    \hline
    $\Omega_b h^2$ & $0.0223$ & $0.02235^{+0.000178}_{-0.000179}$ & $0.0325$ & $0.03945^{+0.0109}_{-0.0145}$ & $0.02226$ & $0.02236^{+0.000168}_{-0.00017}$ & $0.02245$ & $0.02233^{+0.000154}_{-0.000176}$  \\
    $\Omega_c h^2$ & $0.1209$ & $0.124^{+0.0025}_{-0.0039}$ & $0.1257$ & $0.1276^{+0.0087}_{-0.00886}$ & $0.1247$ & $0.1232^{+0.00255}_{-0.00344}$ & $0.1208$ & $0.1207^{+0.00134}_{-0.00116}$  \\
    $100\theta_{MC}$ & $1.041$ & $1.041^{+0.000375}_{-0.000373}$ & $1.031$ & $1.012^{+0.0356}_{-0.0397}$ & $1.041$ & $1.041^{+0.000364}_{-0.000365}$ & $1.041$ & $1.041^{+0.0003}_{-0.000299}$  \\
    $\tau$ & $0.084$ & $0.07043^{+0.018}_{-0.0176}$ & - & - & $0.0409$ & $0.0425^{+0.0162}_{-0.0165}$ & $0.05067$ & $0.04659^{+0.0155}_{-0.016}$  \\
    $\ln (10^{10}A_s)$ & $3.1$ & $3.073^{+0.0351}_{-0.0344}$ & - & - & $3.011$ & $3.013^{+0.0312}_{-0.0318}$ & $3.024$ & $3.022^{+0.0298}_{-0.0305}$  \\
    $n_s$ & $0.9634$ & $0.9609^{+0.00512}_{-0.00518}$ & - & - & $0.9573$ & $0.9615^{+0.00481}_{-0.00479}$ & $0.9645$ & $0.9633^{+0.00406}_{-0.0041}$  \\
    $w$ & $-1.674$ & $-1.691^{+0.318}_{-0.359}$ & $-1.08$ & $-1.172^{+0.171}_{-0.0389}$ & $-1.058$ & $-1.077^{+0.0378}_{-0.0188}$ & $-1.051$ & $-1.07^{+0.0284}_{-0.0163}$  \\
    $\lambda_1$ & $0.0007646$ & $0.001416^{+0.000365}_{-0.00142}$ & $0.005467$ & $0.005257^{+0.00474}_{-0.00526}$ & $0.001596$ & $0.00153^{+0.000574}_{-0.00127}$ & $0.001205$ & $0.0007587^{+0.000335}_{-0.000602}$  \\
    \hline
    $H_0$ & $87.25$ & $84.63^{+15.4}_{-4.9}$ & $68.14$ & $68.33^{+1.}_{-0.979}$ & $65.64$ & $66.89^{+2.47}_{-2.}$ & $67.59$ & $68.4^{+0.807}_{-0.819}$  \\
    $\Omega_{de}$ & $0.8111$ & $0.7859^{+0.07}_{-0.0275}$ & $0.6578$ & $0.6406^{+0.0401}_{-0.0351}$ & $0.6573$ & $0.6717^{+0.0343}_{-0.0214}$ & $0.6851$ & $0.6926^{+0.00921}_{-0.0092}$  \\
    $\Omega_m$ & $0.1889$ & $0.2141^{+0.0275}_{-0.07}$ & $0.3422$ & $0.3594^{+0.0351}_{-0.0401}$ & $0.3427$ & $0.3283^{+0.0214}_{-0.0343}$ & $0.3149$ & $0.3074^{+0.0092}_{-0.00921}$  \\
    $\sigma_8$ & $1.006$ & $0.9833^{+0.102}_{-0.0636}$ & $0.7174$ & $0.7079^{+0.035}_{-0.039}$ & $0.7995$ & $0.8024^{+0.018}_{-0.0177}$ & $0.8012$ & $0.8129^{+0.0125}_{-0.0125}$  \\
    Age/Gyr & $13.61$ & $13.71^{+0.102}_{-0.176}$ & $13.54$ & $13.43^{+0.261}_{-0.266}$ & $13.95$ & $13.92^{+0.0808}_{-0.13}$ & $13.87$ & $13.84^{+0.0376}_{-0.0454}$
\end{tabular}
}
\bigskip
\resizebox{\textwidth}{!}{
\begin{tabular}{ccccc}
\hline
$\chi^2_{min} =$    & $\chi^2_{CMB} + \chi^2_{prior}$ & $\chi^2_{BAO} + \chi^2_{SNIa} + \chi^2_{H_0} + \chi^2_{RSD}$ & $\chi^2_{CMB} + \chi^2_{prior} + \chi^2_{RSD}$ & $\chi^2_{CMB} + \chi^2_{prior} + \chi^2_{BAO} + \chi^2_{SNIa} + \chi^2_{H_0} + \chi^2_{RSD}$ \\
\cline{2-5}
     & 12931.37 + 11.90 &  11.18 + 695.79 + 0.56 + 7.23 & 12941.16 + 11.10 + 23.33 & 12947.69 + 11.07 + 23.38 + 695.49 + 0.85 + 17.58 \\
\hline
\end{tabular}
}
\end{table*}

It is worth to mention here another result observed in Tables \ref{tab.model1}-\ref{tab.model4_RSD}. Apart from Model I, where the constraints from Planck alone are large enough, the constrained values of $H_0$ and $\sigma_8$ using Planck data alone are significantly higher than that using any other data set at low redshift. This indicates that there is a tension between the constrained $H_0$ and $\sigma_8$ from Planck and by using low redshift measurements in the interacting dark energy scenario. This tension also exists in the equation of state of DE as was reported in the above discussion.

\section{Conclusions}
\label{sec:conclusions}
Dark energy and dark matter occupy the largest parts of energy contents of the universe. It is natural to consider the interaction between them. In this work we confront the commonly studied phenomenological interactions between dark energy and dark matter to complementary updated observational data sets. We find that although the Planck 2015 data sets have been improved compared with Planck 2013 results, its constraints on Models I and II are not sufficiently improved. This probably is due to the degeneracies in theoretical model parameters. With the Planck 2015 measurements, the constraints on the model parameters for Model III and IV can be clearly improved. Some differences in the fitting results compared with \cite{Costa:2013sva} have been observed, especially for Model III and IV when employing the combination of data sets from the low redshift measurements and Planck 2015 results.

Including the RSD observational data and combing it with other data sets is very interesting. We can rule out the Model I since the previous negative interaction becomes basically negligible when the RSD data are considered. Furthermore for Model II, we are sure that a positive interaction always exists which can contribute to alleviate the coincidence problem. For Model III and IV, although the coupling can be small positive, we are still not confident to rule out the null interaction possibility. However the interesting thing in these two models is that we can argue that the dark energy cannot be cosmological constant there. These interesting results can be attributed to the complementary RSD data we have included in the analyses. We expect that future more precise RSD data can help us to draw more solid conclusions on examining the theoretical models.

In conclusion, with the newly released complementary data sets, we show that the coupled dark energy and dark matter models are viable options to describe the current universe. Moreover these models can help to solve the coincidence problem that the so called concordance $\Lambda$CDM model is suffering. Further examining the interaction between dark energy and dark matter can help us understand better on the nature of dark sectors of the universe.

\acknowledgments
A. C. acknowledges FAPESP and CAPES for the financial support under grant number 2013/26496-2, S\~ao Paulo Research Foundation (FAPESP). X. X. is supported by the South African Research Chairs Initiative of the Department of Science and Technology and National Research Foundation of South Africa as well as the Competitive Programme for Rated Researchers (Grant Number 91552). B. W. would acknowledge financial supports from National Basic Research Program of China (973 Program 2013CB834900) and National Natural Science Foundation of China. E. A. thanks FAPESP and CNPq for the financial support.
 
This work has made use of the computing facilities of the Laboratory of Astroinformatics (IAG/USP, NAT/Unicsul), whose purchase was made possible by the Brazilian agency FAPESP (grant 2009/54006-4) and the INCT-A.

\bibliographystyle{JHEP}
\bibliography{references}

\end{document}